\UseRawInputEncoding
\documentclass{aastex631}

\usepackage{color}
\usepackage[normalem]{ulem}

\usepackage{multirow}
\usepackage{graphicx}
\shorttitle{High-resolution observation with ALMA in N159W-North}
\shortauthors{Tokuda et al.}
\graphicspath{{./}{figures/}}
\begin{document}

\title{An ALMA study of the massive molecular clump N159W-North in the Large Magellanic Cloud:\\ A possible gas flow penetrating one of the most massive protocluster systems in the Local Group}

%\correspondingauthor{August Muench}
%\email{greg.schwarz@aas.org, gus.muench@aas.org}

\author[0000-0002-2062-1600]{Kazuki Tokuda}
\affiliation{Department of Earth and Planetary Sciences, Faculty of Sciences, Kyushu University, Nishi-ku, Fukuoka 819-0395, Japan}
\affiliation{National Astronomical Observatory of Japan, National Institutes of Natural Sciences, 2-21-1 Osawa, Mitaka, Tokyo 181-8588, Japan}
\affiliation{Department of Physics, Graduate School of Science, Osaka Metropolitan University, 1-1 Gakuen-cho, Naka-ku, Sakai, Osaka 599-8531, Japan}

\author[0000-0002-0138-8348]{Taisei Minami}
\affiliation{Department of Physics, Graduate School of Science, Osaka Metropolitan University, 1-1 Gakuen-cho, Naka-ku, Sakai, Osaka 599-8531, Japan}

\author[0000-0002-8966-9856]{Yasuo Fukui}
\affiliation{Department of Physics, Nagoya University, Furo-cho, Chikusa-ku, Nagoya 464-8601, Japan}

\author[0000-0002-7935-8771]{Tsuyoshi Inoue}
\affiliation{Department of Physics, Konan University, Okamoto 8-9-1, Kobe, Japan}

\author{Takeru Nishioka}
\affiliation{Department of Physics, Nagoya University, Furo-cho, Chikusa-ku, Nagoya 464-8601, Japan}

\author[0000-0002-2794-4840]{Kisetsu Tsuge}
\affiliation{Dr. Karl Remeis Observatory and ECAP, Universit$\ddot{a}$t Erlangen-N$\ddot{u}$rnberg, Sternwartstrasse 7, 96049, Bamberg, Germany}

\author[0000-0001-6149-1278]{Sarolta Zahorecz}
\affiliation{National Astronomical Observatory of Japan, National Institutes of Natural Sciences, 2-21-1 Osawa, Mitaka, Tokyo 181-8588, Japan}

\author[0000-0003-2062-5692]{Hidetoshi Sano}
\affiliation{Faculty of Engineering, Gifu University, 1-1 Yanagido, Gifu 501-1193, Japan}
\affiliation{National Astronomical Observatory of Japan, National Institutes of Natural Sciences, 2-21-1 Osawa, Mitaka, Tokyo 181-8588, Japan}

\author[0000-0002-4098-8100]{Ayu Konishi}
\affiliation{Department of Physics, Graduate School of Science, Osaka Metropolitan University, 1-1 Gakuen-cho, Naka-ku, Sakai, Osaka 599-8531, Japan}

\author[0000-0002-3925-9365]{C.-H. Rosie Chen}
\affiliation{Max Planck Institute for Radio Astronomy, Auf dem Huegel 69, D-53121 Bonn, Germany}

\author[0000-0003-2248-6032]{Marta Sewi{\l}o}
\affiliation{CRESST II and Exoplanets and Stellar Astrophysics Laboratory, NASA Goddard Space Flight Center, Greenbelt, MD 20771, USA}
\affiliation{Department of Astronomy, University of Maryland, College Park, MD 20742, USA}
\affiliation{Center for Research and Exploration in Space Science and Technology, NASA Goddard Space Flight Center, Greenbelt, MD 20771, USA}

\author[0000-0003-3229-2899]{Suzanne C. Madden}
\affiliation{AIM, CEA, CNRS, Universit\'e Paris-Saclay, Universit\'e Paris Diderot, Sorbonne Paris Cit\'e, F-91191 Gif-sur-Yvette, France}

\author[0000-0001-6576-6339]{Omnarayani Nayak}
\affiliation{Space Telescope Science Institute, Baltimore, MD 21218, USA}

\author[0000-0003-1549-6435]{Kazuya Saigo}
\affiliation{Department of Physics and Astronomy, Graduate School of Science and Engineering, Kagoshima University, 1-21-35 Korimoto, Kagoshima, Kagoshima 890-0065, Japan}

\author[0000-0003-0732-2937]{Atsushi Nishimura}
\affiliation{Nobeyama Radio Observatory, National Astronomical Observatory of Japan (NAOJ), National Institutes of Natural Sciences (NINS), 462-2 Nobeyama, Minamimaki, Minamisaku, Nagano 384-1305, Japan}

\author[0000-0002-6907-0926]{Kei E. I. Tanaka}
\affiliation{Center for Astrophysics and Space Astronomy, University of Colorado Boulder, Boulder, CO 80309, USA}
\affiliation{National Astronomical Observatory of Japan, National Institutes of Natural Sciences, 2-21-1 Osawa, Mitaka, Tokyo 181-8588, Japan}

\author[0000-0002-0588-5595]{Tsuyoshi Sawada}
\affiliation{National Astronomical Observatory of Japan, National Institutes of Natural Sciences, 2-21-1 Osawa, Mitaka, Tokyo 181-8588, Japan}
\affiliation{Joint ALMA Observatory, Alonso de C\'ordova 3107, Vitacura, Santiago 763-0355, Chile}

\author[0000-0002-4663-6827]{Remy Indebetouw}
\affiliation{Department of Astronomy, University of Virginia, PO Box 400325, Charlottesville, VA 22904, USA}
\affiliation{National Radio Astronomy Observatory, 520 Edgemont Rd, Charlottesville, VA 22903, USA}

\author[0000-0002-1411-5410]{Kengo Tachihara}
\affiliation{Department of Physics, Nagoya University, Furo-cho, Chikusa-ku, Nagoya 464-8601, Japan}

\author[0000-0001-7813-0380]{Akiko Kawamura}
\affiliation{National Astronomical Observatory of Japan, National Institutes of Natural Sciences, 2-21-1 Osawa, Mitaka, Tokyo 181-8588, Japan}

\author[0000-0001-7826-3837]{Toshikazu Onishi}
\affiliation{Department of Physics, Graduate School of Science, Osaka Metropolitan University, 1-1 Gakuen-cho, Naka-ku, Sakai, Osaka 599-8531, Japan}

\begin{abstract}
Massive dense clumps in the Large Magellanic Cloud can be an important laboratory to explore the formation of populous clusters. We report multiscale ALMA observations of the N159W-North clump, which is the most CO-intense region in the galaxy. High-resolution CO isotope and 1.3\,mm continuum observations with an angular resolution of $\sim$0\farcs25($\sim$0.07\,pc) revealed more than five protostellar sources with CO outflows within the main ridge clump. 
One of the thermal continuum sources, MMS-2, shows especially massive/dense nature whose total H$_2$ mass and peak column density are $\sim$10$^{4}$\,$M_{\odot}$ and $\sim$10$^{24}$\,cm$^{-2}$, respectively, and harbors massive ($\sim$100\,$M_{\odot}$) starless core candidates identified as its internal substructures. The main ridge containing this source can be categorized as one of the most massive protocluster systems in the Local Group. The CO high-resolution observations found several distinct filamentary clouds extending southward from the star-forming spots. The CO (1-0) data set with a larger field of view reveals a conical-shaped, $\sim$30\,pc long complex extending toward the northern direction. These features indicate that a large-scale gas compression event may have produced the massive star-forming complex. Based on the striking similarity between the N159W-North complex and the previously reported other two high-mass star-forming clouds in the nearby regions, we propose a $``$teardrops inflow model$"$ that explains the synchronized, extreme star formation across $>$50\,pc, including one of the most massive protocluster clumps in the Local Group.
\end{abstract}


\section{Introduction}\label{sec:intro}

High-mass stars and the associated events in their formation and evolution dramatically change the surrounding interstellar environment and eventually regulate galaxy evolution. Its astrophysical importance has motivated a large number of both observational and theoretical efforts in the past few decades (see reviews by e.g., \citealt{Zinnecker_2007,Tan_2014}). One of the basic ideas to explain the formation of an individual high-mass star or binary/multiple systems is that a massive cloud core, whose mass is $\sim$100\,$M_{\odot}$, collapses into stellar objects in a similar manner to low-mass star formation \citep[e.g.,][]{McKee_2003}. The dynamical process produces more massive protostellar disks and more energetic outflow than those in the low-mass case \citep[e.g.,][]{Matsushita_2017}, which is consistent with recent observational findings achieved by millimeter/submillimeter facilities \citep[e.g.,][]{Beuther_2002,Hirota_2017,Matsushita_2019,Motogi_2019,TanakaK_2020,Torii_2021}.

Because most stars form as clusters \citep[e.g.,][]{Lada_2003}, understanding the cluster formation mechanism is a key toward establishing the conclusive picture of high-mass star formation. The 24 most massive giant molecular clouds (GMCs) are currently responsible for most of the star formation in the Galaxy \citep{Lee_2012}, assuming the GMCs will make clusters. Observational characterizations of such massive cradles forming young massive clusters (YMC) whose total stellar mass exceeds $\sim$10$^{4}$\,$M_{\odot}$ \citep{Portegies_2010} are thus a reasonable approach to investigate their formation origin. According to the review by \citet{Longmore_2014}, $``$in-situ star formation$"$ requires sufficient gas to be packed into the final cluster volume, and thus predicts the formation of a starless massive clump before the onset of extreme star formation. The mass and size (radius) requirements are $\gtrsim$10$^{4}$\,$M_{\odot}$ and $\lesssim$1\,pc, respectively. Although the probability of discovering a completely starless phase is very small due to the extremely short lifetime (\citealt{Motte_2018,Vazquez_2019}, and references therein), there should be a sufficient number of ongoing star-forming clouds \citep{Ginsburg_2012}. In order to form such massive clouds, mass accumulation, i.e., gas flow, from a large-scale surrounding environment is necessary, and some promising mechanisms have been proposed, mainly motivated by observations, such as cloud-cloud collision \citep{Fukui_2021}, and global hierarchical collapse \citep{Motte_2018}. Revealing a high-dynamic-range molecular gas spatial/velocity distribution from a giant molecular cloud to an individual core inside the parental cloud is still a key method to further consider what mass accumulation mechanism is efficiently working at each physical scale.

The most nearby cluster-forming regions, such as Orion, whose distance from the Sun is $\lesssim$1\,kpc, are not necessarily appropriate to pursue the YMC formation judging from the current gas distribution and star formation activities therein \citep{Portegies_2010}. In this regard, several outstanding regions along with the Galactic disk, e.g., Cygnus~X, Carina, W43, W49, and W51, are the nearest neighbors as extreme star-forming environments, because they have a sufficient molecular gas mass, $\gtrsim$10$^{6}$\,$M_{\odot}$ \citep[e.g.,][]{Nguyen_2016}. 
Unfortunately, almost all of these regions are located within a $\pm$1$\arcdeg$ galactic latitude range on the inner Galactic plane where line-of-sight contaminations from unrelated objects are significant. Because dense clumps/cores rarely appear by chance in the same observed direction, source confusion may not have serious impact on such a small-scale study itself. However confusion from larger ($\gtrsim$10\,pc) scale foreground/background emission to individual massive clumps makes it difficult to examine the formation mechanisms of progenitors leading to YMCs in these outstanding Galactic regions.
Extragalactic targets at a distance of $\lesssim$100\,kpc can be a frontier to obtain a much clearer view around YMC progenitor clumps. Atacama Large Millimeter/submillimeter Array (ALMA) provides an unprecedented view resolving individual proto/prestellar sources with a spatial resolution of $\lesssim$0.1\,pc \cite[e.g.,][]{Indebetouw_2020}.

The Large Magellanic Cloud (LMC) is one of the ideal laboratories to investigate molecular clouds clearly thanks to its proximity ($D\sim$50\,kpc, \citealt{Schaefer_2008,de_Grijs_2014}) and face-on view \citep{Balbinot_2015}. The star formation activity is different from that in the Galaxy; populous clusters are actively forming \citep[e.g.,][]{Hodge_1961,van_1981}. Their stellar masses are 10$^4$--10$^5$\,$M_{\odot}$, which is larger than those of Galactic open clusters \citep{Kumai93,Hunter03}. The most extreme star-forming region is 30~Dor, which is the brightest H$\;${\sc ii} region in the galaxy, inhering a total stellar mass of $\sim$10$^{5}$\,$M_{\odot}$ \citep[e.g.,][]{Cignoni_2015}. However, the total amount of the remaining molecular material in this region is $\sim$10$^{5}$\,$M_{\odot}$ \citep[e.g.,][]{Minamidani_2008,Minamidani_2011,Wong_2011}, which is not large compared to the above-listed GMCs in the Galaxy. Among the $\sim$300 entities discovered by the NANTEN CO(1--0) survey in the LMC \citep{Fukui_2008,Kawamura_2009}, the molecular cloud associated with the N159 H$\;${\sc ii} region (hereafter, the N159~GMC) is the most massive one with a total gas mass of $\sim$2\,$\times$10$^{6}$\,$M_{\odot}$ and a less evolved region than 30~Dor (see also \citealt{Johansson_1994,1998A&A...331..857J}). 

Using the single-dish Atacama Submillimeter Telescope Experiment (ASTE), \citet{Minamidani_2008} performed CO(3--2) observations on several GMCs along the molecular ridge region at the south-east side of the galaxy and confirmed that the N159W cloud shows the strongest intensity. The subsequent comprehensive CO(1--0) study with the Mopra telescope \citep{Wong_2011} also confirmed that it shows the strongest peak along with the N113 region. Using the same telescope, \cite{Nishimura_2016} reported marginal detections of N$_2$H$^{+}$ emission in N159W and N113 despite of a significant deficient of N-bearing molecules in LMC \citep{Millar_1990}. Measurements at millimeter/submillimeter windows \citep{Paron_2016,Galametz_2020} detected several high-density gas tracers, HCO$^{+}$, HCN, CS, and (tentative) C$^{18}$O, toward the N159W region. \cite{Ott2010} performed an NH$_3$ survey on such CO prominent regions, including N159W and N113, using the Australia Telescope Compact Array (ATCA). The NH$_3$ emission was only found in the N159W-North region, indicating the molecule is also deficient in the LMC, and likely traces the densest best-shielded cores. The CO peak and the position of NH$_3$ core is $\sim$10\,pc displaced from the H$\alpha$ peak \cite[e.g.,][]{Smith_1999}, indicating that the molecular gas dissipation has not been started yet. Nevertheless, the Spitzer-based studies discovered several luminous ($\gtrsim$10$^{5}$\,$L_{\odot}$) young stellar object (YSO) candidates inside the cloud \citep[e.g.,][]{Seale_2009,Seale_2014,Gruendl_2009,Chen_2010,Carlson_2012,Jones_2017}. 
These observational signatures strongly demonstrate that active star formation is ongoing in the massive cloud and its evolutionary stage must be early.

\begin{figure*}[htbp]
\centering
\includegraphics[width=120mm]{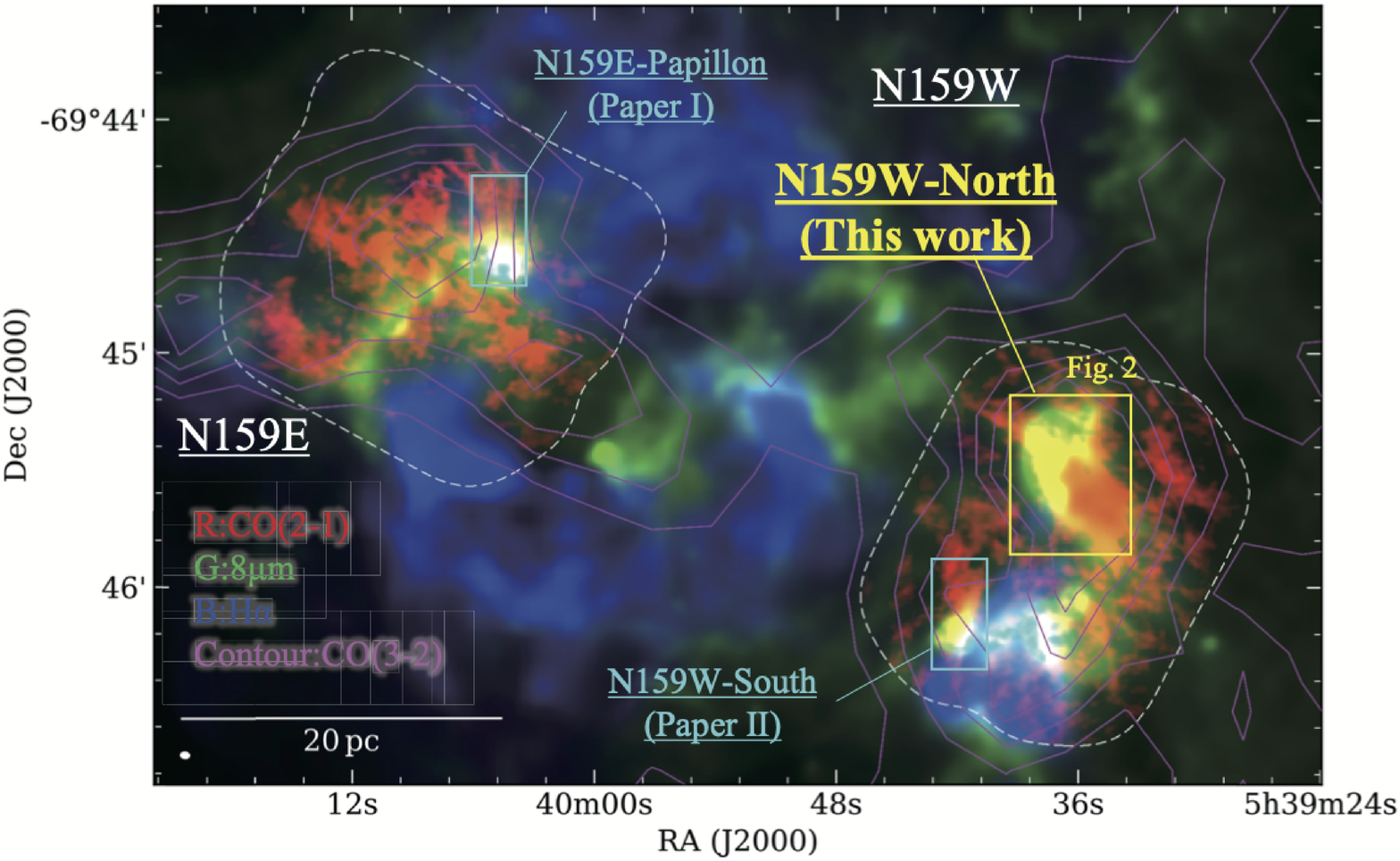}
\caption{A multi-wavelength overview of N159E/W. (a) Three-color composite image combining the ALMA Cycle~1 $^{12}$CO ($J$ = 2--1) (red; \citealt{Fukui_2015,Saigo_2017,Nayak_2018}), the Spitzer/IRAC 8.0\,$\mu$m (green; \citealt{Margaret_2006}), and the MCELS H$\alpha$ (blue; \citealt{Smith_1999}) images. Regions delimited by white dashed lines represent the ALMA Cycle~1 field coverages. The white ellipse in the lower left corner gives the ALMA beam size of 1\farcs3 $\times$ 0\farcs8. The magenta contours show the $^{12}$CO($J$ = 3--2) integrated intensity map with ASTE \citep{Minamidani_2008}. The lowest and subsequent contour steps are 10\,K\,km\,s$^{-1}$. 
\label{fig:N159EW_Cy1_4}}
\end{figure*}

Early ALMA observations in Cycle~1 clearly revealed molecular filaments in the N159W and N159E regions \citep{Fukui_2015,Saigo_2017}. Figure~\ref{fig:N159EW_Cy1_4} shows an overview of the ALMA CO observations in N159E/W. One of the interesting findings is that the locations of the bright YSOs are the intersections of multiple filamentary clouds, which partially resemble galactic hub filaments \citep[e.g.,][]{Myers_2009,Peretto_2013}. The N159 studies hypothesized that filament-filament collisions promoted high-mass star formation activity \citep{Fukui_2015,Saigo_2017}. The follow-up ALMA high-resolution studies in Cycle~4 toward N159E-Papillion (\citealt{Fukui_2019}, hereafter Paper~I; \citealt{Piyush_2021}) and N159W-South (\citealt{Tokuda_2019}, hereafter Paper~II) further resolved complex web-type systems with a filament widths of $\sim$0.1\,pc (Figure~\ref{fig:N159EW_Cy1_4}b,c). Papers~I and II concluded that the coalescence/collision of many individual filaments is unlikely to form such a complex system almost simultaneously across 50\,pc. Alternatively, they suggested that a large-scale H$\;${\sc i} gas flow driven by the last tidal interaction between the LMC and Small Magellanic Cloud (SMC) triggered the filament and subsequent high-mass star formation at the two systems in N159E/W.

The present study is the third paper to show the same observing project with Papers~I and II, but we describe the high-resolution ($\sim$0\farcs26) ALMA 1.3\,mm wavelength CO, its isotopes, and continuum images in the N159W-North region for the first time. The total molecular gas mass of star-forming dense clump are one order of magnitude larger than the previously reported two regions (see the comparison in Sect.~\ref{R:C18O}). Taking into account the fact that high-density gas tracers such as NH$_3$ have also been detected, this region is even more favorable for exploring the formation of massive dense clump leading to YMC.
We also present one more additional data set, CO(1--0), at a lower resolution ($\sim$1\farcs5) covering a wide field coverage. Section~\ref{sec:observe} explains the observations and data reduction. Section~\ref{sec:results} provides a multiscale molecular cloud view from $\sim$100\,pc down to $\sim$0.06\,pc and characterizes the individual proto/prestellar cores. Section~\ref{sec:discuss} discusses the formation scenario of the massive clump in the N159W-North region as well as the high-mass star formation throughout the N159
E/W region.

\section{Observations and data reduction} \label{sec:observe}

In the ALMA Cycle~4, we carried out Band~6 (211--275\,GHz) observations toward the N159W-North region with the ALMA 12\,m array as part of a multiobjects survey in N159 (P.I.: Y. Fukui \#2016.1.01173.S). Because the previous studies (Papers~I and II) described the observation and data reduction details, we briefly summarize the data quality here. The frequency setting mainly targeted the molecular lines of $^{12}$CO($J$ = 2--1), $^{13}$CO($J$ = 2--1), C$^{18}$O($J$ = 2--1) and 1.3\,mm continuum. The synthesized beam sizes of the $^{12}$CO line and continuum images are 0\farcs27 $\times$ 0\farcs23, and 0\farcs26 $\times$0\farcs23, respectively. The rms noise (1$\sigma$) levels of the $^{12}$CO line and the continuum are $\sim$3.7\,mJy\,beam$^{-1}$ ($\sim$1.3\,K) at a velocity resolution of 0.2\,km\,s$^{-1}$ and $\sim$0.027\,mJy\,beam$^{-1}$. Table~\ref{tab:rms} in Appendix~\ref{A:obsbeam} presents the beam propeties and sensitivities of the analyzed lines in this paper. Because the $^{12}$CO and $^{13}$CO emission of the N159W-North region in the Cycle~4 data have non negligible missing flux ($\sim$40--60\%), we used the combined data with the new Cycle~4 and the previously obtained Cycle~1 data sets \citep{Fukui_2015,Nayak_2018} throughout the manuscript. Note that the previously obtained CO sets did not include the compact array (7\,m+TP) data, but the 12\,m array data alone sufficiently covers the total flux judging from single-dish measurements in this region \citep[see][]{Fukui_2015,Nayak_2018}.

In addition to the Band~6 data, we retrieved an ALMA archival Band~3 (84--116\,GHz) CO($J$ = 1--0) data (P.I.: T. Sawada \#2019.1.00915.S) of the same source covering a much larger spatial scale. The field coverage was 170$\arcsec$ $\times$ 180$\arcsec$ centered at ($\alpha_{\rm J2000.0}$, $\delta_{\rm J2000.0}$) = (5$^{\rm h}$39$^{\rm m}$35\fs0,$-$69\arcdeg45\farcs31\farcs2). Using CASA (Common Astronomy Software Application) version 5.6.1, we individually imaged the 12\,m, 7\,m , and TP(Total Power) array data and then combined them together using the \texttt{feather} task. The resultant beam size and sensitivity of the CO($J$ = 1--0) data cube are 2\farcs2 $\times$ 1\farcs8 and $\sim$0.6\,mJy\,beam$^{-1}$ ($\sim$1.5\,K) at a velocity resolution of 0.2\,km\,s$^{-1}$. Note that we made all of the molecular line cube data in a unified velocity frame of LSRK (Local Standard of Rest, Kinematic), and describe the velocity in the same frame throughout the manuscript.

%%%%%%%%%%%%%%%%%%%%%%%%%%%%%%%%%%%%%%%%%%%%%%%%%%%%%%%%%%%%%%%%%%%%%%%%%%%

\section{Results} \label{sec:results}

\subsection{High-resolution $^{12}$CO and $^{13}$CO images of the N159W-North region}\label{R:Cy4CO}

\begin{figure*}[htbp]
\centering
\includegraphics[width=180mm]{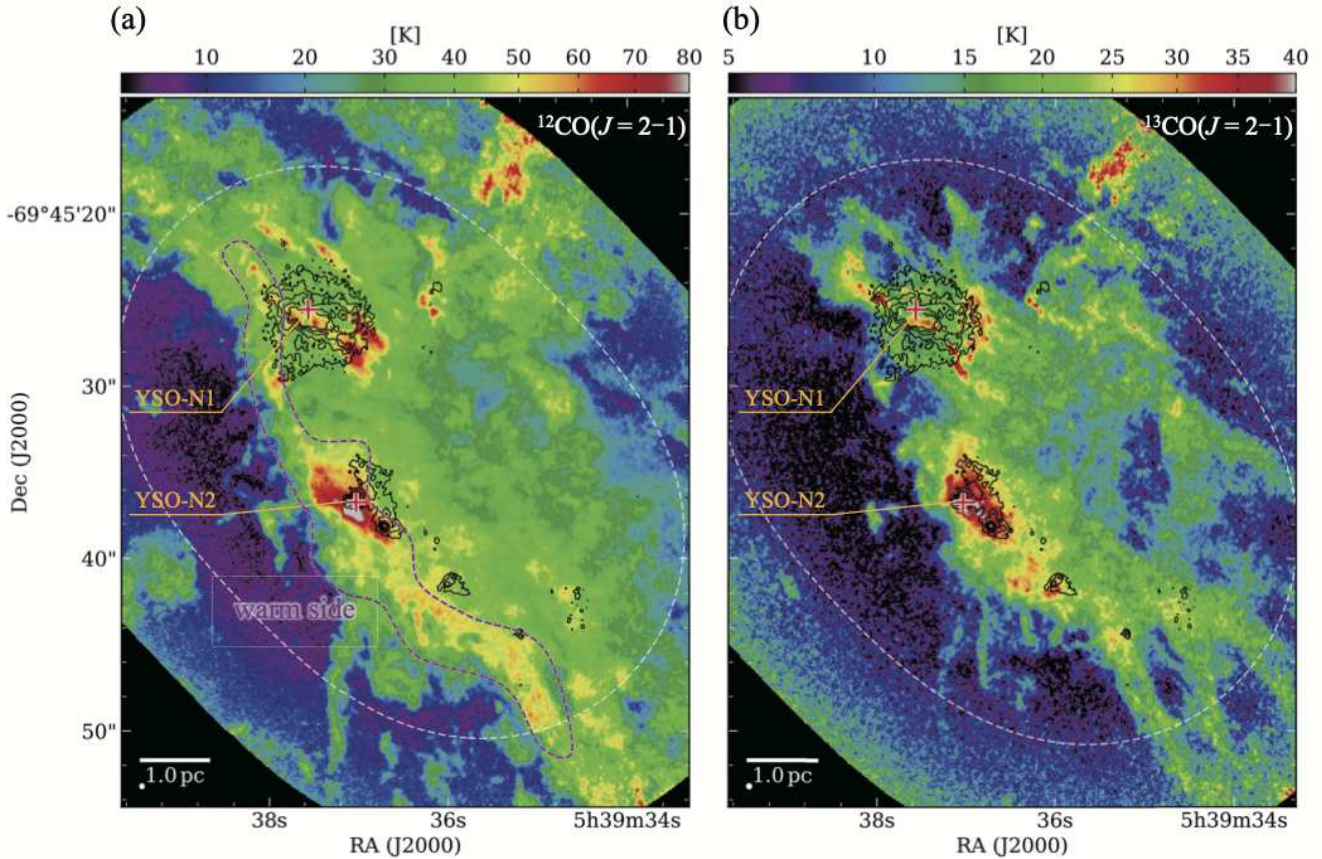}
\caption{
(a) The color-scale image illustrates the $^{12}$CO($J$ = 2--1) peak brightness temperature map. The white ellipse in the lower-left corner gives the angular resolution, 0\farcs27 $\times$ 0\farcs23.
The white dashed ellipse indicates where the mosaic sensitivity falls to 50\%. The purple dashed line marks the warm side of the cloud (see details in Sect.~\ref{D:warmedge}). The black contours show the 1.3\,mm continuum image. The lowest contour and subsequent steps are 0.1\,mJy\,beam$^{-1}$. The red crosses denote positions of two YSOs \citep{Chen_2010}. (b) Same as (a) but for the $^{13}$CO($J$ = 2--1) peak brightness temperature map.
\label{fig:Cy4moms}}
\end{figure*}

Figure~\ref{fig:Cy4moms} shows peak brightness temperature maps of $^{12}$CO and $^{13}$CO($J$ = 2--1) at a spatial resolution of $\sim$0.06\,pc. The high-spatial dynamic range observations have revealed the molecular cloud distribution with a size scale of $\sim$10\,pc down to $\lesssim$0.1\,pc (the velocity-channel maps are presented in Figures~\ref{fig:12COchan} and \ref{fig:13COchan} of Appendix~\ref{A:Cy4chanmap}). Within the molecular cloud, there are at least two spectroscopically confirmed YSOs, whose luminosity are $\sim$10$^{5}$\,$L_{\odot}$ identified by the infrared observations using Spitzer \citep[e.g.,][]{Seale_2009,Gruendl_2009,Chen_2010,Seale_2014,Jones_2017}. In this paper, we call the two sources as YSO-N1 and YSO-N2, which are 053937.56-694525.4 and 053937.04-694536.7, respectively, in the \cite{Chen_2010} catalog. 
For the two YSOs, we detected 1.3\,mm continuum emission (see black contours in Figure~\ref{fig:Cy4moms}), tracing free-free emission from ionized gas and/or thermal dust components in the cold/dense condensations (see more details in Sect.~\ref{R:mm_outflow}). With respect to the continuum and YSO sources, the $^{12}$CO and $^{13}$CO cloud asymmetrically distributes in the northwest(NW)-southeast(SE) direction. We see a highly extended emission on the NW side, while the SE region has a sharp boundary at the cloud edge.  

The $^{12}$CO peak brightness temperature map (Figure~\ref{fig:Cy4moms}a) shows that most of the emitting area exceed 30\,K and the maximum value is $\sim$80\,K. Assuming the abundance (isotope) ratio, $^{12}$CO/$^{13}$CO of 50 \citep{Johansson_1994,Mizuno__2010}, the typical intensity ratio between $^{12}$CO and $^{13}$CO of $\sim$2 tells us that the optical thickness of $^{12}$CO is $\gtrsim$30, which is an optically thick regime. The peak brightness map in $^{12}$CO thus reflects the temperature distribution in this region under the local thermodynamical equilibrium (LTE) assumption. The most intense region is around YSO-N2, but the area near the cloud edge at the SE side is also uniformly high, $\sim$50\,K. In addition to the feedback from YSO-N1, there may be some heating mechanisms changing the temperature of the molecular cloud across several parsecs. We call the distinct feature $``$warm side$"$ hereafter as indicated in the dotted line in Figure~\ref{fig:Cy4moms}a and discuss the possible origins in Sect.~\ref{D:warmedge}.

$^{13}$CO emission is optically thin at a molecular cloud scale generally, and thus it is a good tracer of column density. The LTE calculation \citep{Rohlfs_2004} tells us that the median and maximum H$_2$ column densities are $\sim$3 $\times$10$^{23}$\,cm$^{-2}$ and $\sim$1 $\times$10$^{24}$\,cm$^{-2}$, respectively, assuming [H$_2$]/[$^{13}$CO] = 3${\times}$10$^{6}$ for consistency with the previous analysis \citep[e.g.,][]{Mizuno__2010}.
Adopting the mean molecular weight per hydrogen molecule (2.8), the total H$_2$ gas mass is $\sim$7$\times$\,10$^4$\,$M_{\odot}$ within the $^{13}$CO detected region above 5$\sigma$ noise level on the integrated intensity image. This mass is almost comparable to a single giant molecular cloud traced by $^{13}$CO($J$ = 2--1) (e.g., the Orion~A cloud; \citealt{Nishimura__2015}).

\subsubsection{Filaments extending from the main ridge}\label{R:subfil}

\begin{figure*}[htbp]
\centering
\includegraphics[width=120mm]{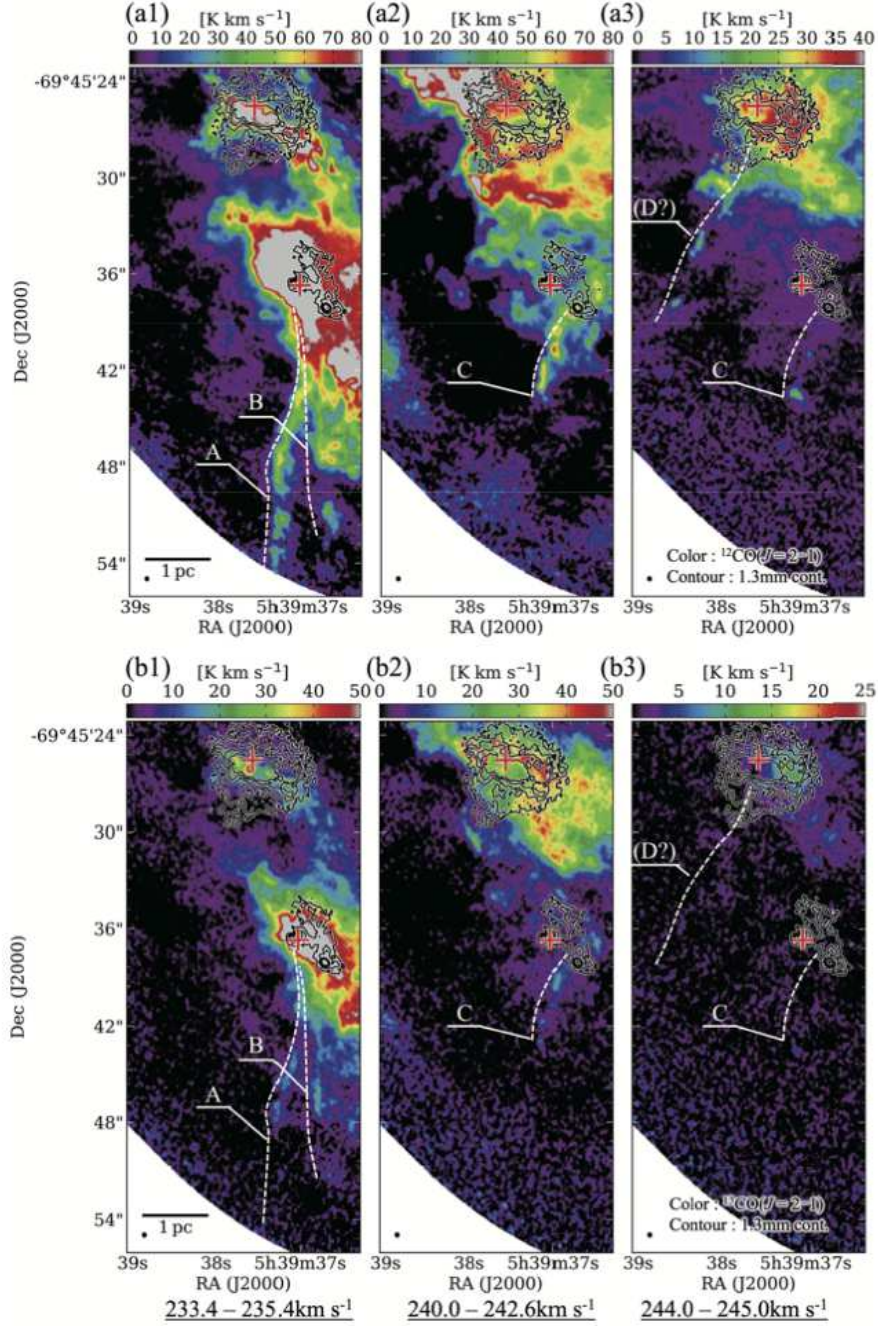}
\caption{Spatial distributions of the filaments in the N159W-North region. (a1--a3) The color-scale images illustrate the velocity channel maps of the $^{12}$CO($J$ = 2--1) with velocity ranges of 232.0--234.0\,km\,s$^{-1}$ (a1), 240.0--242.6\,km\,s$^{-1}$ (a2), and 244.0--245.0\,km\,s$^{-1}$ (a3). The ellipse at the lower-left corner in each panel gives angular resolution, 0\farcs27 $\times$ 0\farcs23. White dashed lines highlight the filaments by eye. Red crosses denote the positions of YSO, same as those in Figure~\ref{fig:Cy4moms}. (b1--b3) Same as panels (a1--a3), but for the $^{13}$CO($J$ = 2--1) maps.} %(c1--c4) Red and blue profiles show the $^{12}$CO and $^{13}$CO spectra at the positions denoted by cyan crosses in panels (a1--a3, and b1--b3). Note that we arbitrarily chose the positions to avoid the complex velocity components originating from the main ride region.} %Green vertical lines indicate the velocity range of each sub-filament.
\label{fig:filament}
\end{figure*}

One of the notable characteristics in the $^{12}$CO and $^{13}$CO images  (Figure~\ref{fig:filament}) is that there are several filamentary structures, which extend to the south direction from the vicinity of YSO-N1 and N2. We call these features as $``$filaments$"$, hereafter. Only filament D has a discontinuous distribution with weaker intensity than the others and is not detected in $^{13}$CO, and thus we treat it as marginal.
The projected lengths of the $^{12}$CO and $^{13}$CO filaments are $\sim$4\,pc and $\sim$3\,pc, respectively. 
Since we could not measure the actual length of the $^{12}$CO filaments, especially for A and B, due to the limited field coverage (Figure~\ref{fig:filament}), the current estimate is thus the lower limit. %The lower resolution CO($J$ = 1--0) data (Sect.~\ref{R:CO10}) shows a sign of southward extension beyond the Cycle~4 field, but we do not deeply discuss it here to satisfy a consistency among the four filaments.
The column density and line masses of the filaments are $\sim$(2--6)${\times}$10$^{22}$\,cm$^{-2}$ and $\sim$(1--3)$\times$10$^{2}$\,$M_{\odot}$\,pc$^{-1}$, respectively, which are listed in Table~\ref{tab:subfilaments}. 

The velocity-widths, $\Delta v$ are $\sim$1--3\,km\,s$^{-1}$ in FWHM, corresponding to velocity-dispersion, $\sigma_v$ ( = $\Delta v$/2$\sqrt{2\rm ln2}$) of 0.4--1.3\,km\,s$^{-1}$. The virial line masses of the filaments, $M_{\rm line,vir} = 2\sigma_{v}^2/G$, are calculated to be (0.9--8)\,$\times$10$^{2}$\,$M_{\odot}$\,pc$^{-1}$, which is somewhat larger than those values obtained from the $^{12}$CO luminosity mass mentioned above.
%larger than those of the $^{12}$CO luminosity mass, $M_{\rm line}$ of $\sim$(1--3)\,$\times$10$^2$\,$M_{\odot}$\,pc$^{-1}$. 
Sect.~\ref{D:subfil} discuss the formation of these filaments and the relation between them and the star formation in this region.　

\begin{table*}[htbp]
    %\centering
     \caption{Physical quantities of filaments traced in $^{12}$CO($J$ = 2--1)}
    \begin{tabular}{ccccccc} \hline \hline
    filament name & $v_{\rm cent}$ [km\,s$^{-1}$]$^{\rm a}$ & ${\Delta}v$ [km s$^{-1}$]$^{\rm a}$& $L$ [pc]$^{\rm b}$ &  $M$ [$M_{\odot}$]$^{\rm c}$ & $N_{\rm H_2}^{\rm ave}$ [cm$^{-2}$]$^{\rm c}$ & $M_{\rm line}$ [$M_{\odot}$\,pc$^{-1}$]$^{\rm d}$ \\
    \hline 
    A & 232.3 & 2.3 & 3.5 & 4${\times}$10$^{2}$ & 2${\times}$10$^{22}$ & 1${\times}$10$^{2}$  \\
    B & 234.5 & 2.4 & 2.9 & 4${\times}$10$^{2}$ & 2${\times}$10$^{22}$ & 1${\times}$10$^{2}$  \\
    C & 242.4 & 3.1 & 1.9 & 3${\times}$10$^{2}$ & 2${\times}$10$^{22}$ & 1${\times}$10$^{2}$ \\
    D & 244.0 & 1.1 & $\cdots^{\rm e}$ & 2${\times}$10$^{2}$ & 2${\times}$10$^{22}$ & $\cdots^{\rm e}$ \\
    \hline
    \end{tabular}
    %\begin{flushleft} 
    \tablenotetext{\rm a}{Central velocity ($v_{\rm cent}$) and width in FWHM (${\Delta}v$) determined by a single Gaussian fitting to the $^{12}$CO spectra at the cyan cross positions in panels (a,b) of Figure~\ref{fig:filament}. }
    \tablenotetext{\rm b}{Projected length between the southern edge of filaments and the boundary of the main ridge (see also the definition of the main ridge in Sect.~\ref{R:C18O}).}
    \tablenotetext{\rm c}{$^{12}$CO luminosity based total mass and averaged column density of each filament. We assumed a constant conversion factor of $X_{\rm CO}$ = 5 $\times$10$^{20}$\,cm$^{-2}$\,(K\,km\,s$^{-1}$)$^{-1}$ \citep{Hughes_2010} and CO(2--1)/CO(1--0) intensity ratio of 1.0.}
    \tablenotetext{\rm d}{Line mass inferred from $M/L$.}
    \tablenotetext{\rm e}{We did not measure the length because filament~D shows a discontinuous spatial distribution.}
    \label{tab:subfilaments}
\end{table*}

\subsubsection{Velocity structures of the molecular cloud}\label{R:velocity}

\begin{figure*}[htbp]
\centering
\includegraphics[width=130mm]{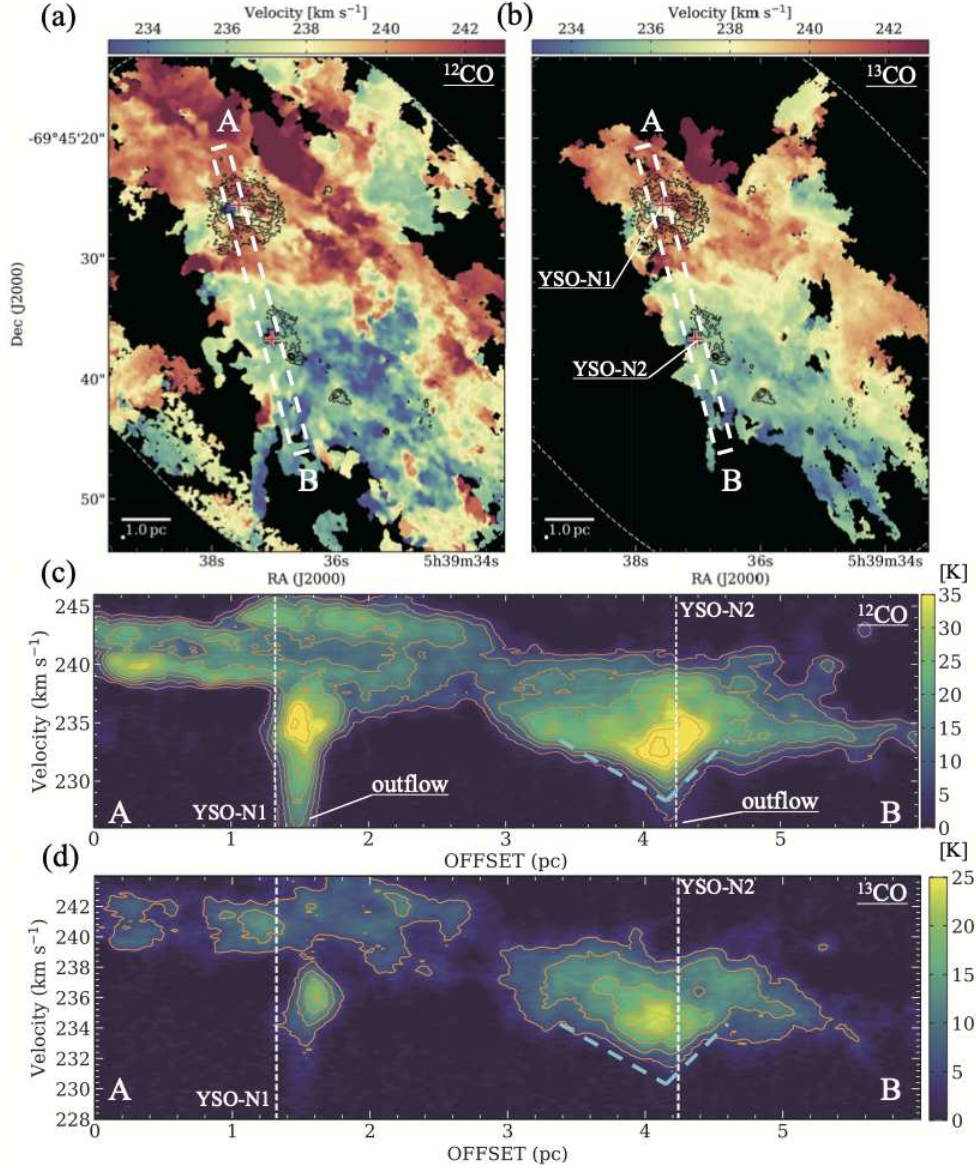}
\caption{
(a) The color-scale image illustrates the first-moment intensity-weighted velocity map of $^{12}$CO($J$ = 2--1). The white ellipse in the lower-left corner gives the angular resolution, 0\farcs27 $\times$ 0\farcs23. The portion of a white dashed ellipse indicates where the mosaic sensitivity falls to 50\%. The rectangle defined by the white thick line marks the extracted region to make the position-velocity diagram in panels (c,d). The black contours show the 1.3\,mm continuum image. The lowest contour and subsequent steps are 0.1\,mJy\,beam$^{-1}$. The red crosses denote positions of two YSOs \citep{Chen_2010}. (b) Same as (a) but for the $^{13}$CO($J$ = 2--1). (c) The color-scale and orange contours show the $^{12}$CO($J$ = 2--1) position-velocity diagram along the rectangle shown in panel (a) The contour levels are [5,10,15,20,25,30,35,40]~K. The two dashed lines represent the positions of YSO-N1 and -N2.
(d) Same as (c) but for $^{13}$CO($J$ = 2--1). The contour levels are [5,10,15,20]~K.
\label{fig:Cy4mom1PV}}
\end{figure*}

Figure~\ref{fig:Cy4mom1PV} shows the velocity structure of $^{12}$CO and $^{13}$CO in the N159W-North region. Overall, the structure is very complex, with a mixture of various components over a velocity range of 10\,km\,s$^{-1}$. One of the outstanding features is that the north side is redshifted, while the south side is blueshifted. This trend is quantitatively similar to the N159E-Papillon and N159W-South regions (Papers I and II). 
We extracted position-velocity (PV) diagrams across the two YSO locations to examine the velocity structure around them in more detail (Figure~\ref{fig:Cy4mom1PV}c,d). The velocity is distributed discontinuously with a boundary around the offset axis of $\sim$3\,pc. As one can see, there are velocity width enhancements toward the two YSO positions. A particular feature seen in $^{12}$CO is the outflow contribution from the YSOs (Sect.~\ref{R:mm_outflow}). 
Although the outflow itself likely contributes only a few $\times$ 0.1 pc, there is also a V-shaped structure around YSO-N2 on the PV diagram over $\sim$1\,pc. This feature may represent a cloud with multiple velocity components interacting with each other (see discussion in Sect.~\ref{D:CCC}).

\subsection{Millimeter continuum sources and molecular outflow}\label{R:mm_outflow}

\begin{figure*}[htbp]
\centering
\includegraphics[width=180mm]{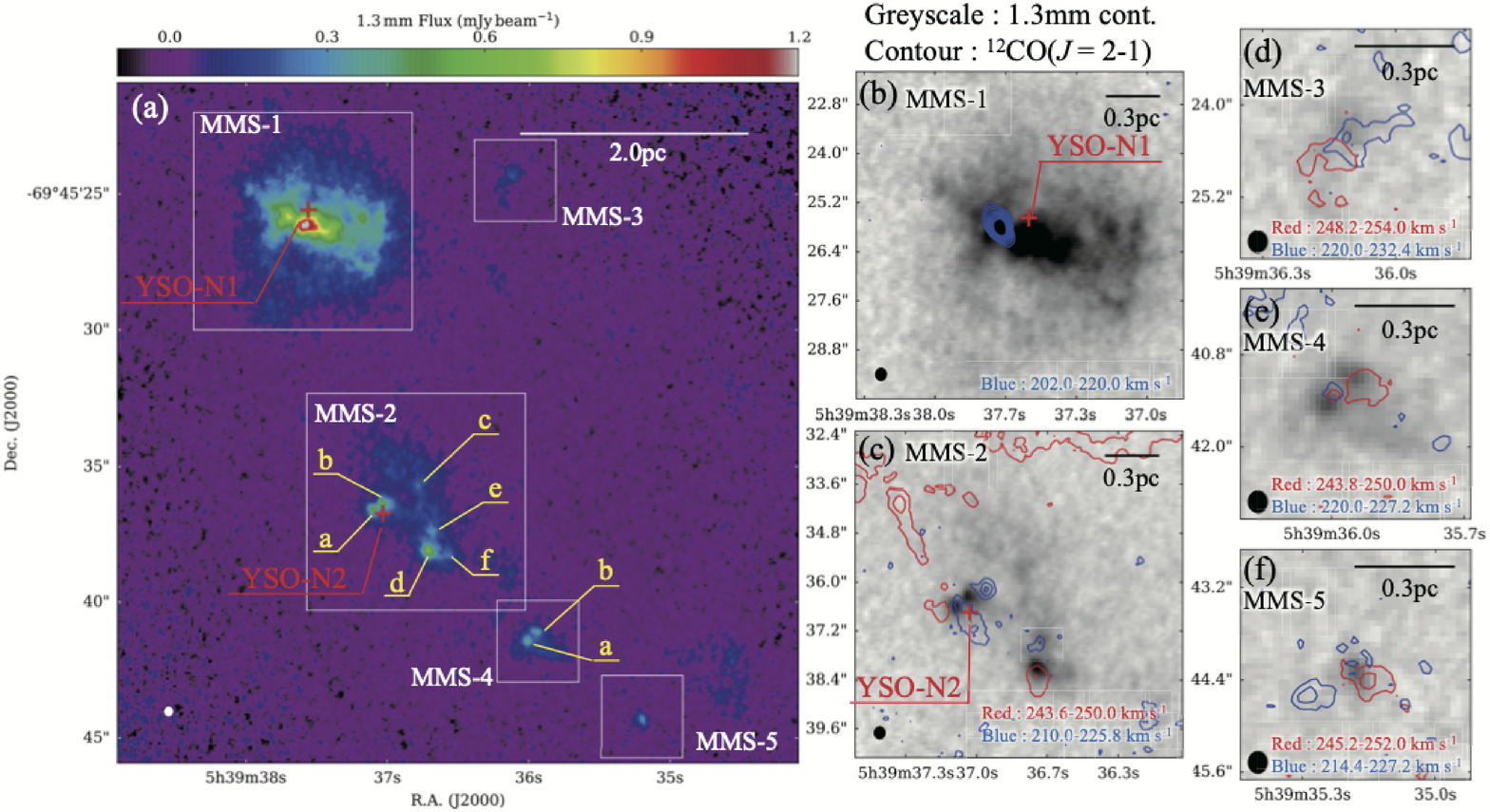}
\caption{1.3\,mm continuum and $^{12}$CO high-velocity emission toward N159W-North. (a) The color-scale image illustrates the 1.3\,mm continuum emission. The white ellipse in the lower-left corner gives the beam size of 0\farcs26 $\times$ 0\farcs23. The red crosses indicate the positions of YSOs \citep{Chen_2010} identified using the Spitzer data with an aperture radius of 3\farcs6. (b--f) Enlarged views of each millimeter source. Red and blue contours shows the redshifted and blueshifted high-velocity emission of $^{12}$CO($J$ = 2--1). The integrated velocity range are given in lower right corners in each panel. We arbitrarily adjusted the contour level to emphasize the weaker components in each target. Black ellipses in each lower left corner show the beam size of 1.3\,mm continuum.
\label{fig:Cy4cont}}
\end{figure*}

Figure~\ref{fig:Cy4cont} shows the 1.3\,mm continuum distributions in the N159W-North region. 
We focus on the stronger than 5$\sigma$ emission within the 3$\sigma$ contour, and we find that there are five spatially well-separated entities. We labeled them as MMS-1, -2, -3, -4, and -5 (see rectangles in Figure~\ref{fig:Cy4cont}). MMS-1--5 have a projected distance of 1--2\,pc from the nearest neighbors. 
The largest/brightest one is MMS-1, coinciding with YSO-N1. \cite{Indebetouw_2004} detected 3\,cm continuum emission toward this source, and our ALMA observations of H30$\alpha$ show a similar spatial extension to the 1.3\,mm emission. 
These results demonstrate that the free-free emission from ionized material largely contribute to the continuum emission from MMS-1. The N159E-Papillon region shows a similar 1.3\,mm continuum structure whose size scale is $\sim$5$\arcsec$ (Paper~I; see also Figure~\ref{fig:C18O}b), but it is visible in the optical H$\alpha$ band, unlike MMS-1 in N159W-North. The Papillon nebula is more evolved than N159W-North~MMS-1 and/or one of the complex velocity components in the foreground hides the ionized gas. In fact, a part of 1.3\,mm emission in MMS-1 overlaps with the C$^{18}$O distribution (see Figure~\ref{fig:MMS2C18O}a). In the other 1.3\,mm sources, the previous measurements, such as longer wavelength radio observations, and hydrogen recombination lines, did not find any indication of ionized material. The thermal dust emission is the main contributor, which is consistent with that of the continuum emission spatially correlates with C$^{18}$O well (see Sect.~\ref{R:C18O}). 

Our visual inspection of MMS-2 and MMS-4 revealed six and two local peaks, respectively (Figure~\ref{fig:Cy4cont}a). We defined their boundary as 50\% intensity in each peak, including MMS-3, and -5, to estimate the physical quantities (Table~\ref{tab:MMS_local}). Although multiple peaks can be also identified in MMS-1, we treat it as a single continuous source in this paper, since a large free-free contribution makes it difficult for us to estimate the physical quantities of thermal component with the 1.3\,mm measurement alone. The flux, $F_{\rm 1.3mm}^{\rm MMS}$, is converted to molecular gas mass, $M_{\rm MMS} = D^2F_{\rm 1.3mm}^{\rm MMS}/R_{\rm dg}\kappa_{\rm 1.3mm}B(T_{\rm d})$, where $D$ is the distance to the source, $R_{\rm dg}$ is the dust-to-gas mass ratio, $\kappa_{\rm 1.3mm}$ is the absorption coefficient per unit dust mass at 1.3\,mm, and $B(T_{\rm d})$ is the Planck function at a dust temperature $T_{\rm d}$. Our assumptions are following; $R_{\rm dg}$ = 3.5\,$\times$10$^{-3}$, and $\kappa_{\rm 1.3mm}$ = 1\,cm$^{2}$\,g$^{-1}$ (see, \citealt{Ossenkopf_1994,Herrera_2013,Gordon_2014}). Assuming that the dust and gas temperatures couple in the dense medium at a core scale \cite[e.g.,][]{Young_2004,Ceccarelli_2007}, we applied the gas temperature determined from ammonia observations (16\,K, \citealt{Ott2010}) as $T_{\rm d}$. As summarized in Table~\ref{tab:MMS_local}, the peak column density and mass of the sources are as high as (0.4--1)${\times}$10$^{24}$\,cm$^{-2}$, and (0.7--3)${\times}$10$^{2}$\,$M_{\odot}$, respectively.

\begin{table*}[htbp]
　　\begin{flushleft} 
     \caption{Physical quantities of 1.3\,mm continuum sources in N159W-North}
    \begin{tabular}{lccccccc} \hline \hline
    Name & ${\alpha}$ (J2000.0)$^{\rm a}$ & ${\delta}$(J2000.0)$^{\rm a}$ & $F_{\rm 1.3mm}^{\rm max}$ [mJy beam$^{-1}$] & $F_{\rm 1.3mm}^{\rm MMS}$ [mJy]$^{\rm b}$  & $N_{\rm peak}$[cm$^{-2}$]$^{\rm c}$ & $M_{\rm MMS}$ [$M_{\odot}$]$^{\rm d}$ & I.R. source\\ \hline
    MMS-1$^{\rm e}$& 05:39:37.55 & -69:45:26.11 & 1.36 & 12.2 & $\cdots^{\rm e}$ & $\cdots^{\rm e}$ & Y\\
    MMS-2a & 05:39:37.09 & -69:45:36.56 & 0.56 & 0.8 & 1${\times}$10$^{24}$ & 1${\times}$10$^{2}$ & Y\\
    MMS-2b & 05:39:37.03 & -69:45:36.40 & 0.53 & 1.1 & 1${\times}$10$^{24}$ & 2${\times}$10$^{2}$ & N\\
    MMS-2c & 05:39:36.78 & -69:45:35.67 & 0.24 & 1.1  & 5${\times}$10$^{23}$ & 2${\times}$10$^{2}$ & N\\
    MMS-2d & 05:39:36.71 & -69:45:38.09 & 0.63 & 2.0 & 1${\times}$10$^{24}$ & 3${\times}$10$^{2}$ & Y\\
    MMS-2e & 05:39:36.68 & -69:45:37.39 & 0.31 & 1.1 & 7${\times}$10$^{23}$ & 2${\times}$10$^{2}$ & N\\
    MMS-2f & 05:39:36.57 & -69:45:38.24 & 0.25 & 0.5  & 5${\times}$10$^{23}$ & 1${\times}$10$^{2}$ & N\\
    MMS-3& 05:39:36.13 & -69:45:24.31 & 0.20 & 0.4 & 4${\times}$10$^{23}$ & 7${\times}$10$^{1}$ & N\\
    MMS-4a & 05:39:36.00 & -69:45:41.42 & 0.39 & 0.7 & 8${\times}$10$^{23}$ & 1${\times}$10$^{2}$ & N\\
    MMS-4b & 05:39:35.94 & -69:45:41.10 & 0.34 & 0.7 & 7${\times}$10$^{23}$& 1${\times}$10$^{2}$  & N\\
    MMS-5& 05:39:35.19 & -69:45:24.31 & 0.32 & 0.4 & 7${\times}$10$^{23}$ & 7${\times}$10$^{1}$ & N\\
    \hline
    \end{tabular}
    \tablenotetext{\rm a}{Peak sky coordinate in 1.3\,mm.}
    \tablenotetext{\rm b}{Flux of the 1.3\,mm continuum emission above $\sim$50\% intensity level from each peak.}
    \tablenotetext{\rm c}{H$_2$ column density at each peak.}
    \tablenotetext{\rm d}{Total gas mass within the same region to measure $F_{\rm 1.3mm}^{\rm MMS}$.}
    \tablenotetext{\rm e}{MMS-1 shows multiple peaks, but we treated it as a single source and did not derive the column denisty and mass, because there is a large free-free contamination (see the text).}
    \label{tab:MMS_local}
    \end{flushleft} 
\end{table*}

We investigated the presence or absence of known infrared counterparts in each continuum source to characterize the star formation activities. The Spitzer-based YSO identifications do not have a sufficient angular resolution to determine which sources, MMS-2a and -2b, are brighter in the mid-infrared wavelength. \cite{Bernard_2016} performed high-resolution near-infrared observations using the VLT (Very Large Telescope) at an angular resolution of $\sim$0\farcs1 with a sensitivity of 21.85\,mag in $K_{\rm s}$. We independently verified the positional accuracy of the VLT detected sources using the GAIA DR2 catalog \citep{Bailer-Jones_2018}. MMS-2a is visible in $K_{\rm s}$ band, while MMS-2b was not detected. MMS-2d is detected in $K_{\rm s}$, but the position is slightly ($\sim$0\farcs1) shifted to the east direction from the continuum peak. According to \cite{Bernard_2016}, the source is categorized as a YSO candidate based on the color-magnitude diagram. Figure~\ref{fig:MMS2C18O}b visualizes the presence/absence of the infrared sources in MMS-2. In the other continuum sources, we could not find any known infrared counterparts. 

Our previous studies (Papers~I and II) discovered molecular outflows in $^{12}$CO at some of the 1.3\,mm continuum sources in the N159E-Papillon and W-South regions. The new ALMA measurements enable us to make a powerful guide map additionally searching for star formation activity in infrared quiescent sources (see also some Galactic studies on infrared dark clouds (IRDC), e.g., \citealt{Kong_2019}). In the same manner, described in Paper~II, we manually searched for high-velocity wing emission more than $\sim$10\,km\,s$^{-1}$ away from the systemic velocity determined by the $^{13}$CO spectra at each continuum source.      
Figures~\ref{fig:Cy4cont}b--f present the spatial distributions of $^{12}$CO high-velocity emission (see also the spectra of Figure~\ref{fig:outflow_spect} in Appendix~\ref{A:outflow}). We conclude that the high-velocity emission is most likely due to molecular outflows from embedded protostellar object(s) inside the continuum sources. The high-resolution $^{12}$CO data reveals the outflow population in this region for the first time, except for the monopolar one at YSO-N1 inside MMS-1, which was already reported in the previous lower resolution ALMA study \citep{Fukui_2015}. The characteristics of the outflows are comparable to those in N159E-Papillon/N159W-South (Papers~I and II); the typical size and maximum velocity of the flows are $\sim$0.2\,pc, and $\sim$20\,km\,s$^{-1}$, respectively. Only one exception is the blueshifted emission in MMS-1 whose maximum velocity is not captured by the current observing setting in the Cycle~4 data. An alternative ALMA project using the CO(3--2) line (K., Tanaka et al. in prep.) detected more than $\sim$60\,km\,s$^{-1}$ emission in this source. 

Among the 1.3\,mm continuum sources, MMS-2c, e, and f do not harbor either infrared nor $^{12}$CO high-velocity emission, indicating that these sources are purely starless at least in the high-mass regime (Figure~\ref{fig:MMS2C18O}b). The inferred gas mass is as massive as $\sim$100\,$M_{\odot}$ (see Table~\ref{tab:MMS_local}). Although some Galactic observations using ALMA discovered similar sources in high-mass star-forming regions \citep[e.g.,][]{Kong_2017,Molet_2019,Zhang_2021}, our finding is the first discovery of starless massive core candidates in extragalactic studies at less than 0.1\,pc resolution.

\subsection{Dense molecular gas traced by C$^{18}$O in N159W-North}\label{R:C18O}
\begin{figure*}[htbp]
\centering
\includegraphics[width=180mm]{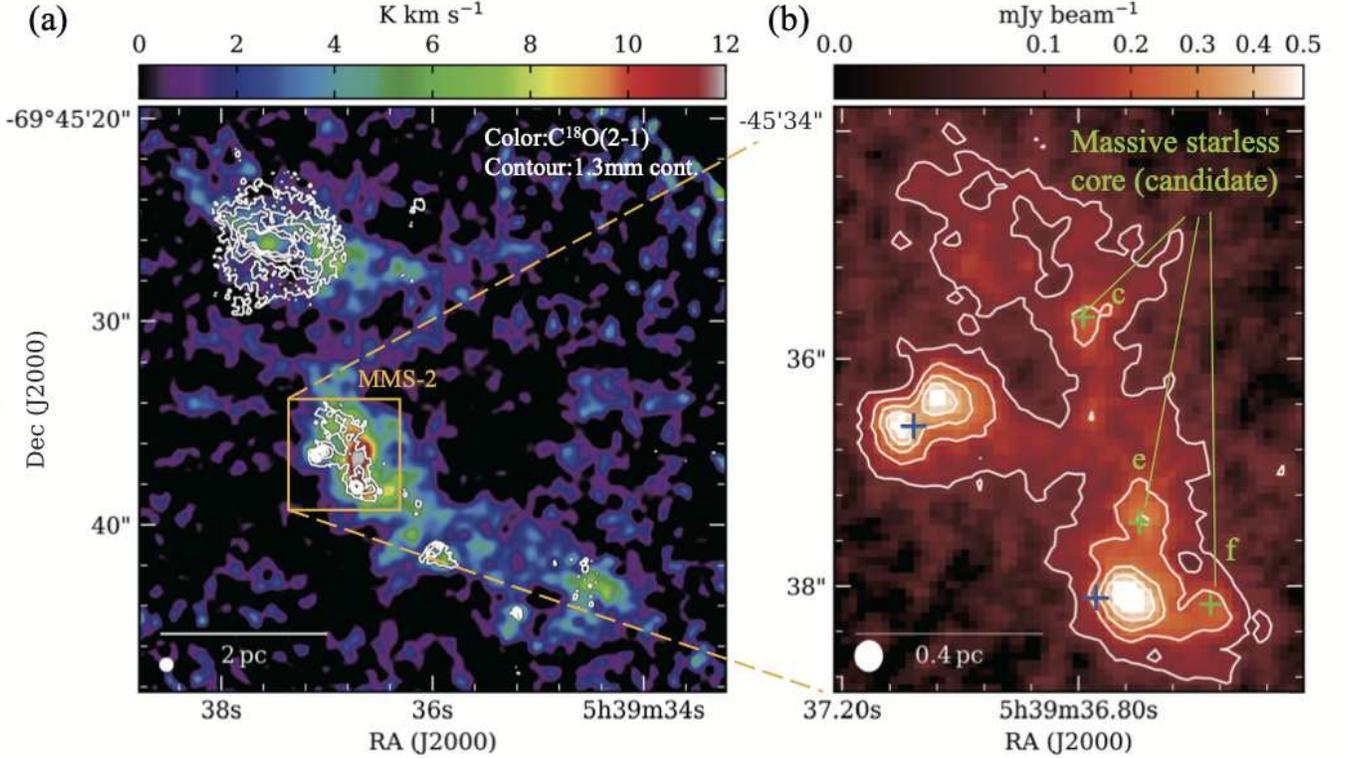}
\caption{C$^{18}$O and 1.3mm continuum emission in the main ridge of N159W-North. (a) The color-scale image illustrates the C$^{18}$O integrated intensity image with an angular resolution of 0\farcs6. The beam size is shown by the white ellipse in the lower left corner. The contours shows the 1.3\,mm continuum emission. The lowest and subsequent contour steps are 0.1\,mJy\,beam$^{-1}$. 
(b) The color-scale image and contours shows an enlarged view of the 1.3\,mm continuum emission toward MMS-2. The contour levels are the same as panel (a). The white ellipse at the lower left corner shows the beam size of 0\farcs26 $\times$ 0\farcs23. The blue and green crosses represent positions of the infrared and starless sources, respectively.} 
\label{fig:MMS2C18O}
\end{figure*}

Figure~\ref{fig:MMS2C18O}a illustrates the C$^{18}$O integrated intensity maps in the N159W-North region. The emitting regions of the rare isotope are more spatially compact than those of $^{12}$CO and $^{13}$CO, and have good spatial correlation with the millimeter sources. These features clearly demonstrate that the C$^{18}$O traces a high density inside of the molecular cloud. 
We defined the dust and gas rich region as $``$main ridge$"$ where we detect both the 1.3\,mm and C$^{18}$O emission and the $^{13}$CO integrated intensity map enables to enclose the dense condensations with a single contour level of $\sim$70\,K\,km\,s$^{-1}$.
%which corresponds to the Main ride as defied in Figure~\ref{fig:CO10_bui}. 
The maximum integrated intensity is $\sim$12\,K\,km\,s$^{-1}$ around MMS-2c, which is the most massive starless source inferred from the 1.3\,mm measurement (Sect.~\ref{R:mm_outflow}). In the MMS-1, the C$^{18}$O emission is relatively weak compared to that in MMS-2, and the extended 1.3\,mm component is not bright in C$^{18}$O. This is because the dense gas around the high-mass protostellar object, YSO-N1, is beginning to dissipate, and the extended 1.3\,mm emission is mainly arising from the ionized gas as described in Sect.~\ref{R:mm_outflow}. 

We quantitively characterize the C$^{18}$O clumps in the N159E/W region (Table~\ref{tab:C18O}, see also Figure~\ref{fig:C18O} in Appendix~\ref{A:C18O}). Our assumption to estimate the virial mass, $M_{\rm vir}$ = 210$R_{\rm clump}\Delta v_{\rm clump}^2$, where $R_{\rm clump}$ is the C$^{18}$O clump radius in parsec and $\Delta v_{\rm clump}$ is the FWHM of the line profile in km\,s$^{-1}$, are a uniform density inside the clump and ignoring the magnetic field contribution. To estimate the luminosity-based H$_2$ mass using the LTE assumption, we need to apply a conversion factor, i.e., relative abundance, [H$_2$]/[C$^{18}$O]. According to Galactic studies \citep[e.g.,][]{Frerking_1982}, a commonly used value of [H$_2$]/[C$^{18}$O] is $\sim$5.9 $\times$10$^{6}$. However, the early single-dish measurements by \cite{Johansson_1994} showed that C$^{18}$O emission in the LMC is quite weak, indicating that [H$_2$]/[C$^{18}$O] in the LMC is about 20 times higher than in the Milky Way (MW), $\sim$1.2 $\times$10$^{8}$. We adopted the higher value as the conversion factor to derive the LTE mass. The resulting masses are one order of magnitude larger than the virial mass (Table~\ref{tab:C18O}), which is typical for such a dense, gravitationally bound system (see also \citealt{Nayak_2018}). The current observational evidence suggests that the main ridge in N159W-North and its subregions have dense gas, whose density and total masses are $\sim$10$^{5}$\,cm$^{-3}$ and $\gtrsim$10$^{4}$\,$M_{\odot}$, respectively.  

\begin{table*}[htbp]
    %\centering
     \caption{Physical properties of C$^{18}$O clumps in N159E/W}
    \begin{tabular}{lccccc} \hline \hline
    Region name  & $R_{\rm clump}$ [pc]$^{\rm a}$ & $\Delta v_{\rm clump}$ [km\,s$^{-1}$]$^{\rm b}$ & $V_{\rm cent}$ [km\,s$^{-1}$]$^{\rm b}$ & $M_{\rm vir}$ [$M_{\odot}$]$^{\rm c}$ & $M_{\rm LTE}$ [$M_{\odot}$]$^{\rm d}$  \\ \hline 
    MMS-1      & 0.63 & 5.9 & 239.9 & 5${\times}$10$^{3}$ & 1${\times}$10$^{4}$ \\  
    MMS-2      & 0.71 & 4.6 & 235.6 & 3${\times}$10$^{3}$ & 2${\times}$10$^{4}$  \\
    MMS-4      & 0.36 & 2.6 & 236.2 & 5${\times}$10$^{2}$ & 1${\times}$10$^{3}$  \\
    MMS-5      & 0.31 & 2.5 & 235.0 & 4${\times}$10$^{2}$ & 1${\times}$10$^{3}$  \\  \hline
    main ridge$^{\rm e}$ & $\cdots$ & $\cdots$ & $\cdots$ & $\cdots$ & 4${\times}$10$^{4}$  \\ \hline  \hline
    N159E-Papillon         & 0.41 & 3.1 & 232.4 & 8${\times}$10$^{2}$ & 7${\times}$10$^{3}$  \\
    N159W-South            & 0.38 & 4.6 & 236.1 & 2${\times}$10$^{3}$ & 5${\times}$10$^{3}$  \\
    \hline
    \end{tabular}
    \tablenotetext{\rm a}{Radius of a circle having the same area above the C$^{18}$O integrated intensity of $\sim$4\,K\,km\,s$^{-1}$, which corresponds to $\sim$5$\sigma$ noise level. }
    \tablenotetext{\rm b}{Velocity width in FWHM ($\Delta v_{\rm clump}$) and central velocity ($V_{\rm cent}$) of the C$^{18}$O spectra at the peak sky coordinate in each clump.}
    \tablenotetext{\rm c}{Virial mass, $M_{\rm vir}$ = 210$R_{\rm clump}\Delta v_{\rm clump}^2$ (see the text).}
    \tablenotetext{\rm d}{The LTE mass assuming the uniform excitation temperature of 20\,K and [H$_2$]/[C$^{18}$O] of $\sim$1.2 $\times$10$^{8}$ (see the text) within the region having $>$5$\sigma$ C$^{18}$O emission.}
    \tablenotetext{\rm e}{We defined only the $M_{\rm LTE}$, because the C$^{18}$O emitting regions are not spatially connected.}
    \label{tab:C18O}
\end{table*}

\subsection{A comprehensive CO view around/in the N159W-North region}\label{R:CO10}

\begin{figure*}[htbp]
\centering
\includegraphics[width=180mm]{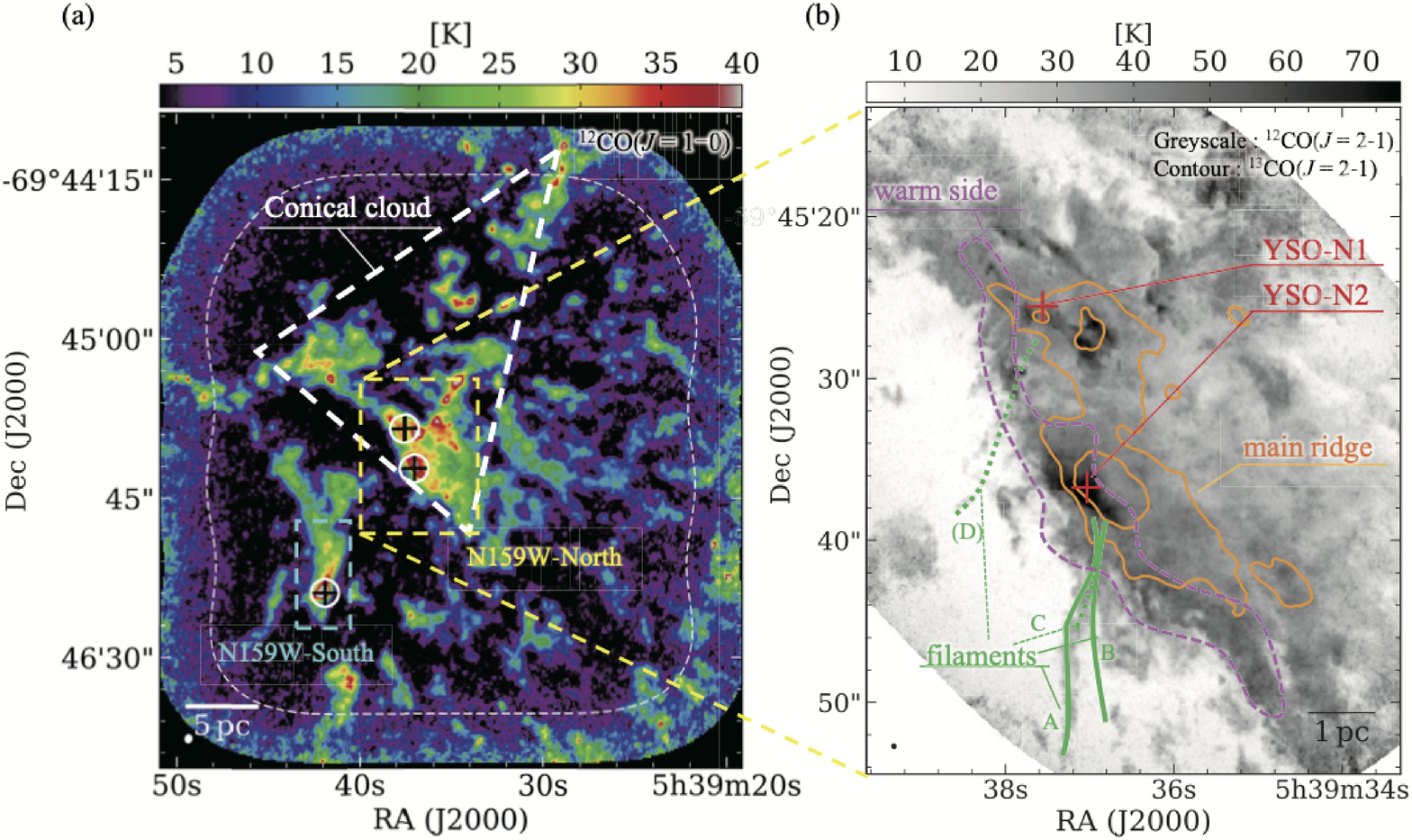}
%\vskip{-0.5cm}
\caption{Multi-spatial-scale molecular gas view around the N159W-North region. (a) The color-scale image illustrates the $^{12}$CO($J$ = 1--0) peak brightness temperature map extracted from a velocity range of 234.7--239.5\,km\,s$^{-1}$. The white dashed contours indicates where the mosaic sensitivity falls to 50\%. The white ellipse in the lower-left corner gives the angular resolution, 2\farcs2 $\times$ 1\farcs8. The region determined by white dashed triangle represent the CO emitting regions defined as the conical cloud. The yellow and cyan dashed rectangles indicate the N159W-North and -South regions studied in this work and Paper~II, respectively. Black crosses with white circles indicate positions of YSO candidates \citep{Chen_2010} within the Cycle~4 observed field. 
(b) An explanatory map of distinct features (see the text) in the N159W-North region on the high-resolution $^{12}$CO($J$ = 2--1) peak brightness temperature image, which is the same as the color-scale image in Figure~\ref{fig:Cy4moms}a. The black ellipse in the lower-left corner gives the angular resolution, 0\farcs27 $\times$ 0\farcs23. Orange contour shows a smoothed $^{13}$CO($J$ = 2--1) velocity-integrated intensity image with contour levels of 70 and 140\,K\,km\,s$^{-1}$.
\label{fig:CO10_bui}}
\end{figure*}

Figure~\ref{fig:CO10_bui}a illustrates the CO($J$ = 1--0) peak brightness temperature map. The field coverage is the widest among the N159W ALMA studies using CO and its isotope lines with a sub-pc resolution. Note that we arbitrarily extracted a velocity range of 234.7--239.5\,km\,s$^{-1}$ to highlight the N159W-North related structures in Figure~\ref{fig:CO10_bui}a. The velocity-channel map is presented in Figure~\ref{fig:CO10chan} of Appendix~\ref{A:Cy7chanmap}.

As one can see, there are at least two elongated or filamentary features intersecting around the main ridge region as indicated by the triangle in Figure~\ref{fig:CO10_bui}a. The total projected lengths of the western and eastern clouds are $\sim$20\,pc and $\sim$30\,pc, respectively. Although their central velocities differ from each other by a few km\,s$^{-1}$, they are spatially connected via the main ridge, making a single large system.
We call this chain of features $``$conical cloud$"$ hereafter. 
These characteristics qualitatively resemble those of the V-shaped filament in the N159W-South region (\citealt{Fukui_2015}; Paper~II).
The southern tip of the conical cloud, which roughly lies to the Cycle~4 field coverage, has CO peak brightness temperature of $\gtrsim$30\,K.
Compared to the entire conical cloud, the temperature rises across the main ridge region, not only the internal subregion, the warm side introduced in Sect.~\ref{R:Cy4CO}.

The total CO(1--0) luminosity of the conical cloud is 3.7${\times}$10$^{3}$\,K\,km\,s$^{-1}$\,pc$^2$, corresponding to $\sim$1.6${\times}$10$^{5}$\,$M_{\odot}$ if we assume a CO-to-H$_2$ conversion factor, 5\,$\times$10$^{20}$\,cm$^{-2}$\,(K\,km\,s$^{-1}$)$^{-1}$ \citep{Hughes_2010}. The conical cloud is an order of magnitude more massive than the N159W-South cloud, $\sim$10$^{4}$\,$M_{\odot}$ (Paper~II).  
Because the comprehensive single-dish survey in the LMC shows that the N159W region is the most CO intense spot (see also Sect.~\ref{sec:intro}), the conical cloud is likely one of the largest and brightest structures in this galaxy. 
Based on the spatial distribution of YSO candidates \citep{Chen_2010}, the western part of the observed field (RA(J2000)$\lesssim$5$^{\rm h}$39$^{\rm m}$30$^{\rm s}$) seems quiescent in high- and intermediate-mass star formation, although we see a lot filamentary structure with a strong intensity contrast. The total molecular gas mass of this quiescent region is $\sim$1.2${\times}$10$^{5}$\,$M_{\odot}$. In summary, about half of the total mass of molecular material is associated with the current high-mass star formation in this observed field.

Figure~\ref{fig:CO10_bui}b highlights the characteristic features presented in Sect.~\ref{R:Cy4CO}, and \ref{R:mm_outflow} to guide the positional relationship with each other on the high-resolution CO map. 

\section{Discussion} \label{sec:discuss}

\subsection{Dynamical status of the characteristic components }\label{D:indv}
This section discusses the physical states of the newly discovered features in N159W-North with our ALMA observations described in Sect.~\ref{sec:results}.

\subsubsection{Filaments traced by $^{12}$CO and $^{13}$CO}\label{D:subfil}
Several filaments are seen extending from the southern edge of the main ridge (Sect.~\ref{R:subfil}). They are relatively ordered in the north-south direction and their roots appear to be spatially attached to the protostellar systems in MMS-1 and MMS-2. 
We discuss here the dynamical state (stability) of the filaments to infer a possible fate. As shown in Sect.~\ref{R:subfil}, the virial line mass of filaments is somewhat larger than those of the $^{12}$CO luminosity mass. If we ignore the stabilizing effect by magnetic force, the filaments are not gravitationally stable objects. In this case, the filaments can be dissipated within a turbulent crossing time, which is an order of $\sim$10$^{5}$\,yr = $W_{\rm fil}$/$\sigma_v$, where $W_{\rm fil}$ is the typical width of filament ($\sim$0.1--0.2\,pc). 
In fact, the current star formation is inactive in all of the filaments except for their junctions with the millimeter sources. 
Among them, filament~D has the weakest and shows the discontinuous $^{12}$CO distribution (Figure~\ref{fig:filament}(a3)), suggesting that this filament is the oldest one and is currently in the process of dissipating. 

We discuss possible formation scenarios of the filaments in the N159W-North region. The filaments spatially connect to the protostellar source, and thus they are likely related to the star formation activity. \cite{Schneider_2010} found similar $^{13}$CO filaments toward the most massive clump in the Galactic Cygnus~X DR21 region. The morphological features and physical properties, such as mass ($M$ $\sim$2600\,$M_{\odot}$) and length ($L$ $\sim$5\,pc), are very similar to N159W-North. The authors interpreted this as filaments falling onto the central ridge due to the global collapse. However, this interpretation is not necessarily supported by gas kinematic features, e.g., accelerated motion toward the star-forming main ridge. A robust common feature of the filaments in the N159W-North regions is that they are distributed only on one side of the main ridge, suggesting the existence of an asymmetric gas flow through the main ridges. The warm side nature can be explained by the past shock wave event in this region (see Sect.~\ref{D:warmedge}), and thus we propose that an extensive gas flow penetrates the N159W-North main ridge and promotes the high-mass star formation with remarkable filaments. We will discuss this possibility further in Sect.~\ref{D:CCC}.

\subsubsection{Warm side}\label{D:warmedge}
 
We found that the peak brightness temperature of $^{12}$CO at the southeast edge of the main ridge is remarkably high ($\gtrsim$50\,K) compared to the typical value ($\sim$30\,K) of this region, indicating that something is contributing to the heating. The probable reasons are (case~1) external radiation field and cosmic ray heating, (case~2a) shock heating in the turbulent material within the cloud, and/or (case~2b) shock heating due to external gas compression. 
\cite{Lee_2016} obtained multi-line CO transitions up to $J$ = 12--11 in the N159W region using Herschel Space Observatory and ground-based telescopes, and estimated the molecular gas kinematic temperature of $>$100\,K and density of $\sim$10$^3$\,cm$^{-3}$ on $\sim$10\,pc scale. Their modeling showed that case~1 is unlikely to produce the warm environment, and instead, suggested that low-velocity C-type shocks (case~2a and/or 2b) with $\sim$10\,km\,s$^{-1}$ appear to be more plausible. 
Our narrow field coverage and high-resolution observations highlight the detection region of the rare isotopes of CO, and we are selectively looking only at denser and colder regions than the \cite{Lee_2016} study.
However, the origin of the cloud nature can still be understood in the same context, because there is no significant external heating source near the ALMA observed region too.
For case~2a, \cite{Tokuda_2018} actually found warm CO (30--60\,K) filamentary components in a turbulent low-mass star-forming dense core in Taurus. However, the physical scale of the gas is quite small as 1000\,au that our present study could not resolve, and shock heating with internal turbulence is expected to realize a more stochastic temperature distribution, instead of asymmetric heating region over several pc as seen in N159W-North. We conclude that mechanical heating due to an external gas colliding flow (case~2b) is probable.

\subsubsection{Dense clumps in the main ridge traced by C$^{18}$O and 1.3\,mm and star-formation activities therein}\label{D:clumps}

We found that the molecular mass of the main ridge of N159W-North and the most massive inside clump exceeds 10$^{4}$\,$M_{\odot}$, despite their compactness with a size of less than a few parsecs. Such a massive clump, which is not frequently discovered in the MW disk region, agrees with the mass and size requirements to evolve into a young massive cluster \citep[e.g.,][]{Longmore_2014}. 
Table~\ref{tab:LocalGroup} lists several massive clumps studied in the Local Group of galaxies (MW, LMC, and M33). Note that a large-scale survey of molecular gas has been conducted at the SMC, but such a massive clump has not been discovered yet \citep{Tokuda_2021}. The mass estimates based on the thermal dust emission and/or molecular line emission are subject to a factor of two or three uncertainties due to methodological differences and systematic errors arising from, e.g., dust properties and molecular abundance. Therefore, we cannot immediately conclude that our characterization of the N159W-North main ridge is the most massive in the Local Group, but it is certainly one of the most massive. The N113 region also has very massive clumps, and thus it could be one of the top contenders in the LMC.
However, because this source was not detected in the ATCA NH$_3$ observations \citep{Ott2010}, the dense gas fraction is likely smaller than that of the N159W-North main ridge.

\begin{table*}[htbp]
    \begin{flushleft} 
     \caption{Physical properties of cluster-forming massive clumps in the Local Group}
    \begin{tabular}{lcccccccc} \hline \hline
    Clump Name & Hosting Galaxy & Size$^{\rm a}$ & $M_{\rm H_2}$ & $N_{\rm H_2}^{\rm peak}$ & $n_{\rm H_2}^{\rm ave}$ $^{\rm b}$ & Beam & Tracers & References$^{\rm c}$  \\
     & & [\,pc] & [M$_{\odot}$] & [cm$^{-2}$] & [cm$^{-3}$] & [pc] & &  \\ \hline
    N159W-N main ridge$^{\rm d}$ & \multirow{2}{*}{LMC} & 1.1$\times$3.3 & $\sim$7$\times$10$^{4}$ & \multirow{2}{*}{$\sim$1$\times$10$^{24}$} & $\sim$1.0$\times$10$^{5}$ & \multirow{2}{*}{$\sim$0.1} & $^{13}$CO/C$^{18}$O/1.3\,mm & \multirow{2}{*}{This work}  \\  
    \hspace{1.43cm} MMS-2 $^{\rm e}$ &  & $\sim$1.5 & $\sim$2$\times$10$^{4}$ &  & $\sim$2$\times$10$^{5}$ &  & C$^{18}$O/1.3\,mm &  \\
    \hline
    N113 Region B   & LMC & 1.2$\times$2.7 & $\sim$2$\times$10$^{4}$ & $\sim$9$\times$10$^{23}$ & $\sim$9$\times$10$^{4}$ & $\sim$0.2& $^{13}$CO/C$^{18}$O/1.3\,mm & [1]  \\
    \hline
    W43 MM1     & \multirow{4}{*}{MW} & 3.9$\times$2.0 & $\sim$2$\times$10$^{4}$ & $\sim$3$\times$10$^{23}$ & $\sim$4$\times$10$^{4}$ & $\sim$0.5 & C$^{18}$O/1.3\,mm & [2,3,4] \\  
    W49-N       &  & $\sim$4 & $\sim$5$\times$10$^{4}$ & $\sim$2$\times$10$^{24}$ & $\sim$2$\times$10$^{4}$ & $\sim$0.1 & $^{13}$CO/C$^{18}$O & [5] \\
    W51 G49.5$-$0.4 &  & $\sim$2.2 & (1--2)$\times$10$^{4}$ & $\sim$4$\times$10$^{23}$ & (3--5)$\times$10$^{4}$ & $\sim$0.5 & C$^{18}$O & [6] \\
    %Carina & MW & & & & & & &  \\
    Cygnus~X DR21 &   & $\sim$4$\times$1 & $\sim$3$\times$10$^{4}$ & $\sim$5$\times$10$^{23}$ & $\sim$10$^{5}$ & $\sim$0.2 & $^{13}$CO/N$_2$H$^{+}$ & [7,8]\\ \hline
    NGC~604 MMS-1 & \multirow{2}{*}{M33} & 2.9$\times$1.9 & $\sim$5$\times$10$^{4}$  & $\cdots$ & $\sim$2$\times$10$^{4}$ & $\sim$2$\times$1 & C$^{18}$O/1.3\,mm & [9]\\
    GMC-16 MMS    &  & 2.2$\times$1.1 & $\sim$2$\times$10$^{4}$ & $\cdots$ & $\sim$2$\times$10$^{4}$ & $\sim$2$\times$1 & C$^{18}$O/1.3\,mm & [10]\\
 \hline \hline
    \end{tabular}
    \tablenotetext{\rm a}{Projected length of major and minor axes or size in diameter.}
    \tablenotetext{\rm b}{Averaged density assuming the uniform, spherical geometry.}
    \tablenotetext{\rm c}{References: [1]We retrieved an ALMA archival data (2015.1.01388, see also \citealt{2018ApJ...853L..19S}) of CO and its isotope lines at 1.3 mm wavelength (Nishioka et al. in prep.) and applied the same mass estimation method as in N159W-North.; [2]\citet{Motte_2003}; [3]\citet{Louvet_2014}; [4]\citet{Kohno_2021}; [5]\citet{2013ApJ...779..121G}; [6]\citet{Fujita_2021}; [7]\citet{Schneider_2010}; [8]\citet{Dobashi_2019}; [9]\citet{Muraoka_2020}; [10]\citet{Tokuda_2020}}
    \tablenotetext{\rm d}{The parameters are determined by the $^{13}$CO intense region as indicated with the orange contours in Figure\,\ref{fig:CO10_bui}b.}
    \tablenotetext{\rm e}{This source is part of the main ridge. The physical properties are the same as those in Tables~\ref{tab:MMS_local} and \ref{tab:C18O}.}
    \label{tab:LocalGroup}
    \end{flushleft} 
\end{table*}

\subsection{Colliding flows promoting the protocluster formation in N159W-North}\label{D:CCC}

We summarize the features described in Sect.~\ref{sec:results} and the possibilities discussed in Sect.~\ref{D:indv}. The cluster forming N159W-North main ridge resides at the apex of a conical-shaped structure traced by the CO(1-0) emission (Sect.~\ref{R:CO10}). The presence of several southward filaments from the main ridge, together with the warm side features, suggests that a north-south penetrating colliding flow may have triggered the formation of a massive clump. 

\begin{figure*}[htbp]
\centering
\includegraphics[width=120mm]{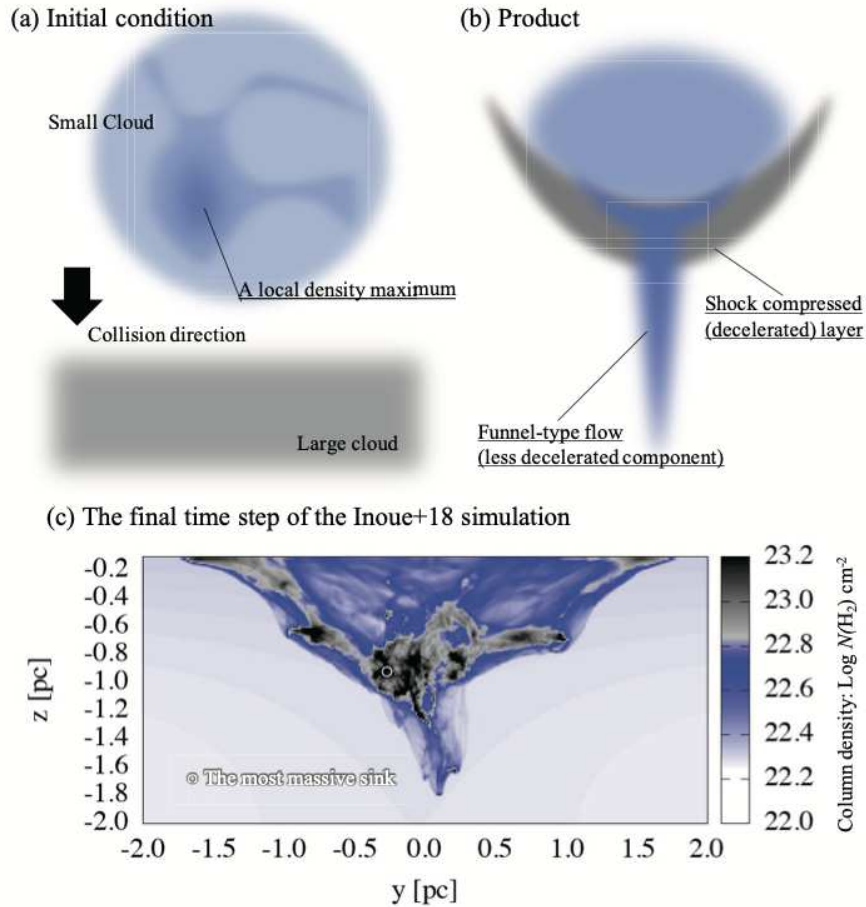}
\caption{Cloud-cloud collision as a possible formation origin of several distinct features, such as the conical cloud, dense cores, and filaments, in the N159W-North region. Panels (a) and (b) schematically illustrate the initial condition and compressed product after the collision, respectively. (c) The final time step of the MHD simulation by \cite{Inoue_2018}.
\label{fig:sim}}
\end{figure*}

We compare these features with a molecular cloud collision model. Many, if not all, of these qualitative characteristics, are explained by the results of numerical simulation of molecular cloud collision by \cite{Inoue_2018}. They simulated that a small cloud collided with a larger cloud with a density of 10$^{3}$\,cm$^{-3}$ at a relative velocity of 10 km\,s$^{-1}$ by solving isothermal magnetohydrodynamic (MHD) equations with self-gravity.
Generally, a gas clump colliding with the larger cloud is decelerated once it interacts or sweeps up the same amount of the larger cloud gas. Since turbulence creates many dense clumps in the small cloud, some dense clumps go beyond the arc-like structure.
Figure~\ref{fig:sim} summarizes such phenomenon schematically, and we call the penetrated filamentary feature $``$funnel-type flow$"$.
In the N159W-North region, the conical cloud and warm side show the shock compression layer, and the main ridge with a mass of $\gtrsim$10$^{4}$\,$M_{\odot}$ is supposed to be the product of the colliding event. If the filaments represent funnel-type flows described above, the feature could also be a piece of indirect evidence of a colliding flow.

We caveat the discrepancy between our observations and the simulation by \cite{Inoue_2018}. The N159W-North system is heavier and larger than that in the simulation. Although it is beyond the scope of this paper to perform an optimized calculation for N159W-North, we expect to see further massive clumps with multiple funnel-type flows, if we perform a colliding flow simulation with a higher mass initial condition, because the hydrodynamic phenomenon is basically scale-free.

We presented the morphological features, temperature distribution, and total mass of the dense clump as indirect evidence for molecular cloud interaction, but the velocity analysis enables us to find additional relevant characteristics (Sect.~\ref{R:velocity}). Paper I shows that the molecular cloud in the N159W-South region has V-shaped structures around protostellar objects on PV diagrams. This feature was originally designed based on numerical simulation reproducing molecular cloud collision phenomena (\citealt{Takahira_2014} and the subsequent synthetic observations in CO emission \citealt{Torii_2017,Fukui_2018a}). In N159W-North, the V-shaped structure is more prominent around YSO-N2, while the same feature around YSO-N2 is less distinct (see Figure~\ref{fig:Cy4mom1PV}) possibly due to the molecular gas ionization (see Sect.~\ref{R:mm_outflow}).

\subsection{High-mass star formation across the N159E/W region: an updated view of the synchronized massive cluster formation}\label{D:HIflow}

We discuss the massive cluster formation throughout the N159E/W regions by combining current knowledge obtained with molecular and atomic gas observations. Paper~I and Paper~II found that the massive protostars in N159E-Papillon and N159W-South, respectively, are accompanied by filamentary molecular clouds with their line mass of 300--1000\,$M_{\odot}$\,pc$^{-1}$. Although these two filamentary systems are separated by about 50\,pc in the sky, their physical and morphological features are quite similar, suggesting the formation of filaments and massive protostars simultaneously on the order of 0.1\,Myr. Papers~I and II proposed that the trigger for these events could be the H$\;${\sc i} colliding flow with a velocity of 50--100\,km\,s$^{-1}$ from the northern direction driven by the last close encounter with the Small Magellanic Cloud $\sim$2\,Gyr ago \citep{Fukui_2017,Tsuge_2019}. 

\begin{figure*}[htbp]
\centering
\includegraphics[width=180mm]{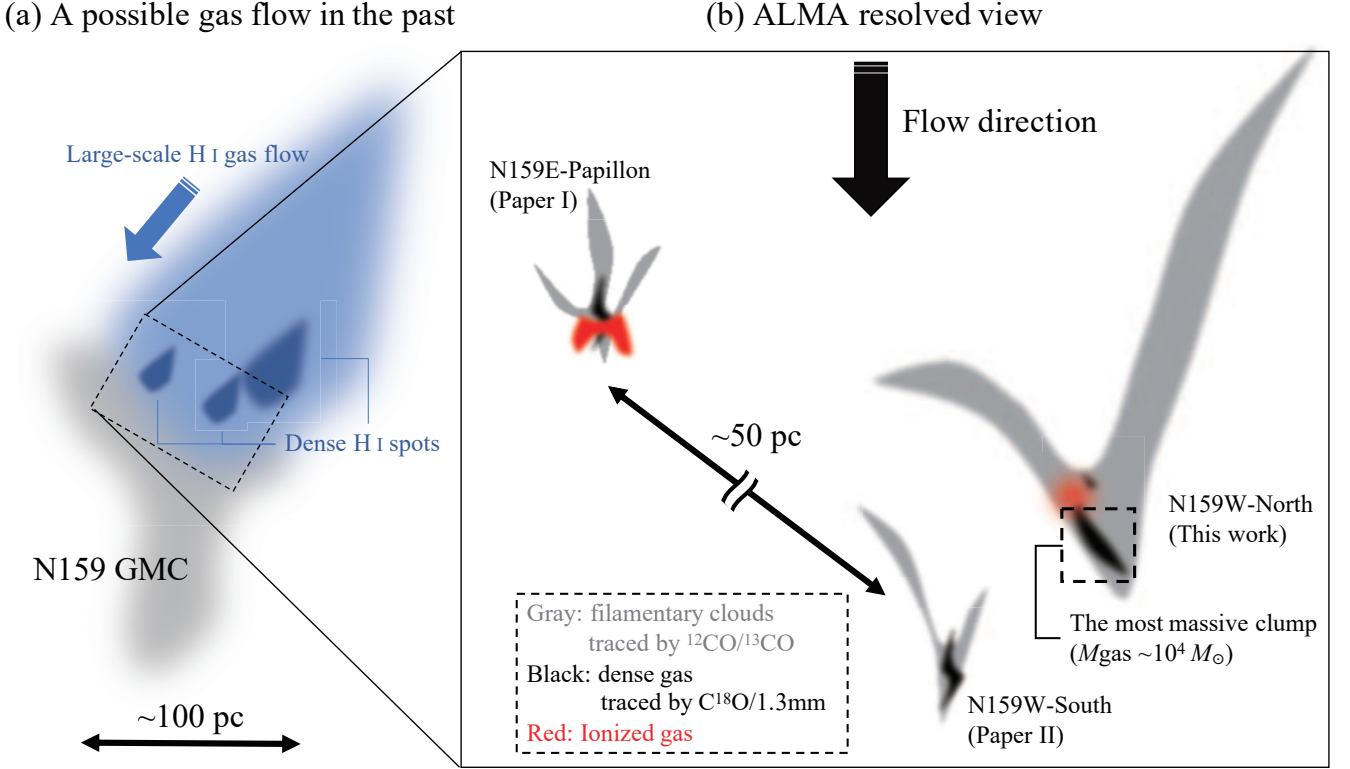}
\caption{Schematic view of the star formation process in N159E/W motivated by the ALMA studies.
\label{fig:N159_view}}
\end{figure*}

Since the newly revealed N159W-North system has a very similar head-tailed filamentary molecular cloud, it is possible that we can interpret its formation scenario in the same context. Figure~\ref{fig:N159_view} shows our updated colliding flow star formation scenario, $``$teardrops inflow model$"$ in the N159 region. In this scenario, the H$\;${\sc i} gas flow containing multiple dense spots interacts with a pre-existing GMC. Locally, this event can be regarded as a collision between large and small clouds, resulting in the morphological and dynamical features described in Sect.~\ref{D:CCC}. N159E-Papillon and N159W-South are impacted by relatively low-mass blobs, while a larger one impacts the N159W-North region, producing a 30\,pc scale compression layer. This scenario potentially explains multiple generations of high-mass star formation of N159E/W. N159E already started the main star formation activities 3-4 Myr ago with its numerous O and early B optically-visible stars, except for the above-mentioned young Papillon region, while N159W just started its main episode with mostly massive YSOs \citep{Chen_2010}. In this line of thought, assuming that the two regions were in the same pre-existing GMC and only N159E was actively star-forming earlier, the H$\;${\sc i} dense blob seems likely to be the main contributor to forming many massive cores, i.e., the YMC progenitor, especially in N159W.

Numerical simulations by \cite{Maeda_2021} demonstrated $\sim$10$^{4}$\,$M_{\odot}$ massive clump formation driven by a large-scale H$\;${\sc i} gas flow that mimics galactic interactions. \cite{Abe_2021} showed that fast ($>$10\,km\,s$^{-1}$) gas flow is more favorable for massive filament formation with a line mass of $>$100\,$M_{\odot}$\,pc$^{-1}$. 
From observational perspectives, recent ALMA studies in M33 also found massive filaments with high-mass cluster formation associated with galaxy-galaxy interactions and other galaxy-scale gas motions related to a spiral arm \citep{Tokuda_2020,Muraoka_2020}. Galactic high-resolution, large-scale molecular gas surveys toward active cluster-forming regions listed in Table~\ref{tab:LocalGroup} exclusively show complex velocity structures originated from multiple-gas components \citep{Miyawaki_2009,Miyawaki_2021,Dobashi_2019,Kohno_2021,Fujita_2021}. The total mass of their parent clouds exceeds 10$^6$\,$M_{\odot}$, which is an order of magnitude higher than the typical GMC mass, and thus the collision/coalescence of clouds and/or the large-scale diffuse gas inflow from the surrounding environment likely contribute to the formation of gravitationally bound massive clumps that leads to YMC formation.

In the N159E-Papillon and N159W-South regions, the total filamentary gas and dense gas masses are $\sim$10$^{4}$\,$M_{\odot}$ and $\sim$10$^{3}$\,$M_{\odot}$, respectively, suggesting they may be just medium-sized clusters. On the other hand, in N159W-North, the total gas mass of the large-scale filamentary cloud (conical cloud) and dense gas (main ridge) exceed 10$^5$\,$M_{\odot}$ and 10$^4$\,$M_{\odot}$, respectively, indicating that the system grows into a more massive cluster. \cite{Fukushima_2021} numerically demonstrated that a 10$^4$\,$M_{\odot}$ molecular cloud packed into a few pc volume is converted into a cluster with a high star formation efficiency because the gravitational potential is stronger than the negative feedback force of the stars formed. 
This case allows forming a massive cluster with a mass of $\gtrsim$10$^4$\,$M_{\odot}$ in the N159W-North region.
Furthermore, since a part of the conical cloud with a total mass of $\sim$10$^{5}$\,$M_{\odot}$ is connected to the main ridge, it may grow into a massive cluster comparable to R136 in 30~Dor if the conveyor belt-like gas supply mechanism where gas accretion and star formation occur simultaneously \citep{Longmore_2014, Krumholz_2020} works in this region. 

Although the large-scale H$\;${\sc i} flow scenario was originally designed to explain the R136 formation \citep{Fukui_2017}, the fact that the same flow distributes over a kpc scale motivates us to apply it describing the star formation history in N159. In a reimportation of this context, an ALMA wide-field imaging study found a N159-type conical-shaped molecular filament system around R136 (T. Wong et al. submitted). These observational pieces of evidence may demonstrate that galaxy-scale (H$\;${\sc i}) gas flows and the subsequent more than a few $\times$ 10\,pc scale filament system development are universally responsible for the super star cluster formation that can drive galaxy evolution itself. Since clusters larger than 10$^5$\,$M_{\odot}$ are only found in a limited number in the Local Group, further searches for similar filament systems in more remote galaxies will help to elucidate the super star cluster formation mechanism in the local universe.

\section{Conclusions} \label{sec:conclusion}

Our ALMA observations revealed a comprehensive molecular gas view of the N159W-North region in the LMC with a high-dynamic spatial range from a few $\times$10\,pc to $\lesssim$0.1\,pc. The main results and conclusions are summarized as follows:

\begin{itemize}

\item The $^{12}$CO and $^{13}$CO high-resolution data depicted highly complex spatial and velocity structures of the N159W-North molecular cloud. The $^{12}$CO brightness temperature is higher than $\sim$30\,K throughout the observed region and even higher than $\sim$50\,K, especially in the southern side of the cloud. One of the outstanding features is that several filaments, whose length and line masses are a few parsecs and $\gtrsim$100\,$M_{\odot}$\,pc$^{-1}$, extend from high-mass star-forming dense cores toward the southern direction.

\item Based on $^{12}$CO high-velocity emission, which appears to be of outflow origin, we found five new bright star-forming cores in 1.3\,mm continuum emission at the infrared quiescent spots. We also identified $\sim$0.1\,pc scale starless cores with a mass of $\sim$100\,$M_{\odot}$, which were not discovered in the previous extragalactic studies. The dense molecular region (main ridge) traced by thermal dust continuum and C$^{18}$O harbors the above-mentioned proto- and prestellar cores. 
The total mass of the cluster-forming clump exceeds 10$^4$\,$M_{\odot}$ despite its compact size of a few pc, making it one of the most massive, dense categories in the Local Group of galaxies.

\item The wide-field CO($J$ = 1--0) map deciphered a head-tailed conical shape feature, whose size scale and total molecular gas mass are $\sim$30\,pc and $\sim$2 $\times$10$^{5}$\,$M_{\odot}$, respectively. The strong impacts of a cloud-cloud collision presumably explain the observed large-scale and small-scale subfeatures, such as the warm side nature and filaments in the southern side of the main ridge.

\item Combining our current observational understanding of two other regions in N159, E-Papillon (Paper~I) and W-South (Paper~II), three systems across more than 50\,pc show active star formation simultaneously.
Massive clumps, especially N159W-North like objects, are highly rare in the Local Group, indicating that a typical galactic environment cannot easily produce YMC precursors. We hypothesize $``$teardrops inflow model$"$ to explain the synchronized, extreme cluster formation, possibly driven by a quite dynamic, substructured flow induced by a galactic-scale phenomenon.
\end{itemize}

\begin{acknowledgments}

We would like to thank the anonymous referee for useful comments that improved the manuscript. This paper makes use of the following ALMA data: ADS/ JAO. ALMA\#2012.1.00554.S, 2016.1.01173.S, and 2019.1.00915.S. ALMA is a partnership of ESO (representing its member states), NSF (USA) and NINS (Japan), together with NRC (Canada), MOST and ASIAA (Taiwan), and KASI (Republic of Korea), in cooperation with the Republic of Chile. The Joint ALMA Observatory is operated by ESO, AUI/NRAO, and NAOJ. This work was supported by NAOJ ALMA Scientific Research grant Nos. 2016-03B, Grants-in-Aid for Scientific Research (KAKENHI) of Japan Society for the Promotion of Science (JSPS; grant Nos. JP18K13582, JP18H05440, JP21H00049, and JP21K13962), and NSF award 2009624. The National Radio Astronomy Observatoryis a facility of the National Science Foundation operatedunder cooperative agreement by Associated Universities,Inc. The material is based upon work supported by NASA under award number 80GSFC21M0002 (M. S.). Dr. Benoit Neichel kindly provided us the VLT photometric data \citep{Bernard_2016} to compare spatial distributions of the ALMA millimeter continuum and infrared sources.
\software{CASA (v5.6.1; \citealt{McMullin07}), Astropy \citep{Astropy18}, APLpy \citep{Robi12}}

\end{acknowledgments}

\appendix \label{App}

\section{Data qualities}\label{A:obsbeam}

Table~\ref{tab:rms} gives the beam sizes and r.m.s. sensitivities of each molecular line data.

\begin{table*}[htbp]
    %\centering
     \caption{Beam Properties and Sensitivities in the molecular line observations}
    \begin{tabular}{lccccc} \hline \hline
    line name & $B_{\rm maj}$ [arcsec] & $B_{\rm min}$ [arcsec] & $B_{\rm P.A.}$ [deg] & r.m.s$_{\rm ch}$ [K]$^{\rm a}$ & r.m.s$_{\rm i.i.}$ [K\,km\,s$^{-1}$]$^{\rm b}$   \\ \hline
    $^{12}$CO($J$ = 2--1) & 0.27 & 0.23 & 6.1 & 1.3 & 1.1  \\
    $^{13}$CO($J$ = 2--1) & 0.28 & 0.24 & 5.6 & 1.4 & 1.1  \\
    C$^{18}$O($J$ = 2--1) & 0.28 & 0.24 & 4.7 & 1.0 & 0.9  \\ \hline
    \end{tabular}
    \tablenotetext{\rm a}{Noise level of data cube at a velocity resolution of 0.2\,km\,s$^{-1}$.}
    \tablenotetext{\rm b}{Noise level of integrated intensity image over a velocity range of 5\,km\,s$^{-1}$, which roughly corresponds to the typical line width of the N159W-North cloud.}
    \label{tab:rms}
\end{table*}

\section{High-resolution CO channel maps}\label{A:Cy4chanmap}

\begin{figure*}[htbp]
\centering
\includegraphics[width=180mm]{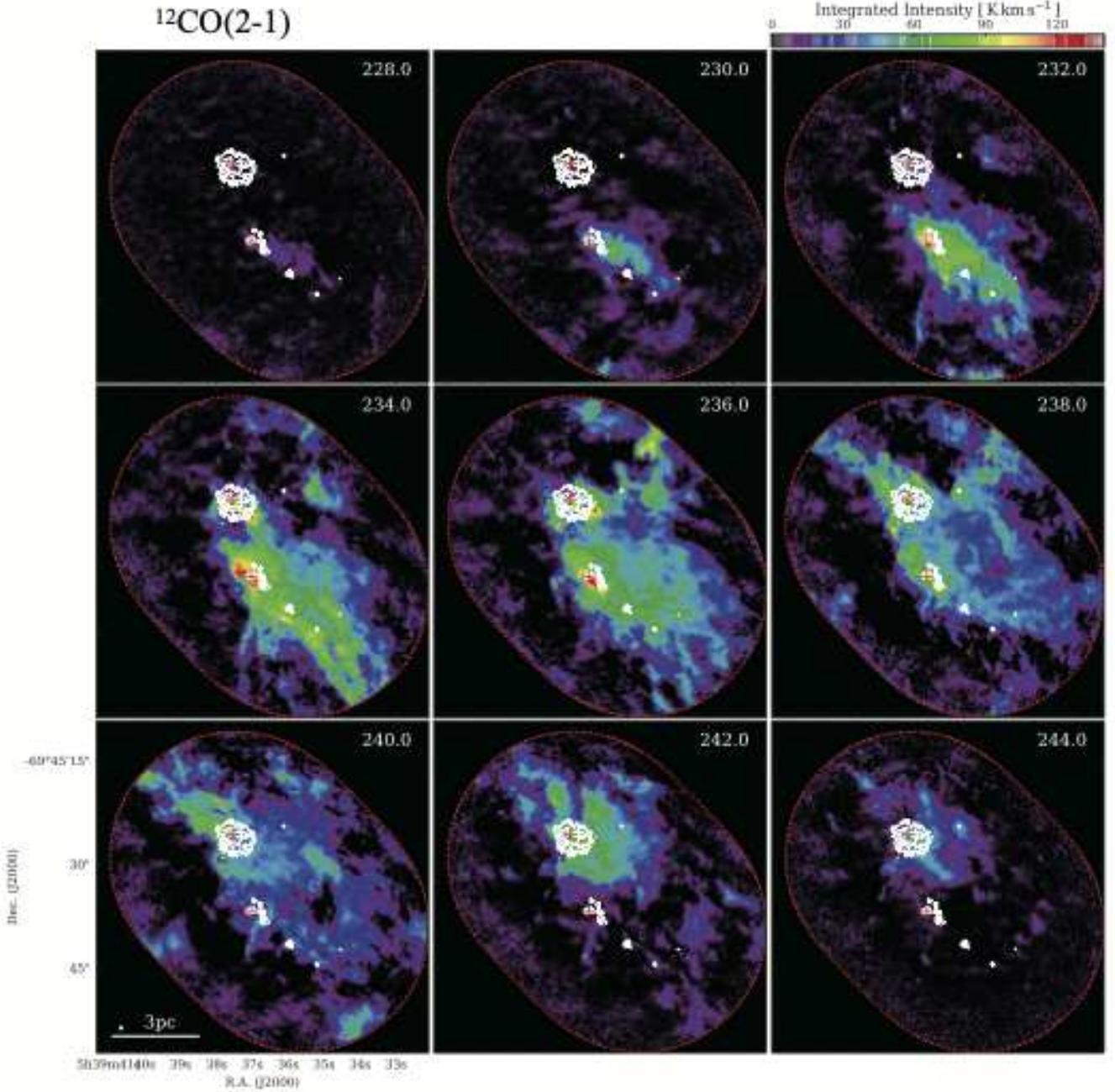}
\caption{Velocity channel maps of N159W-North in $^{12}$CO($J$ = 2--1). The lowest velocity in unit of km\,s$^{-1}$ of each panel are given in the upper right corners. The white ellipse in the lower left corner in the lower left panel shows the beam size, 0\farcs29 $\times$ 0\farcs23, of the $^{12}$CO data. Red dotted lines show the field coverage. The white contours show the 1.3\,mm continuum emission, same as Figure~\ref{fig:Cy4moms}. 
\label{fig:12COchan}}
\end{figure*}

\begin{figure*}[htbp]
\centering
\includegraphics[width=180mm]{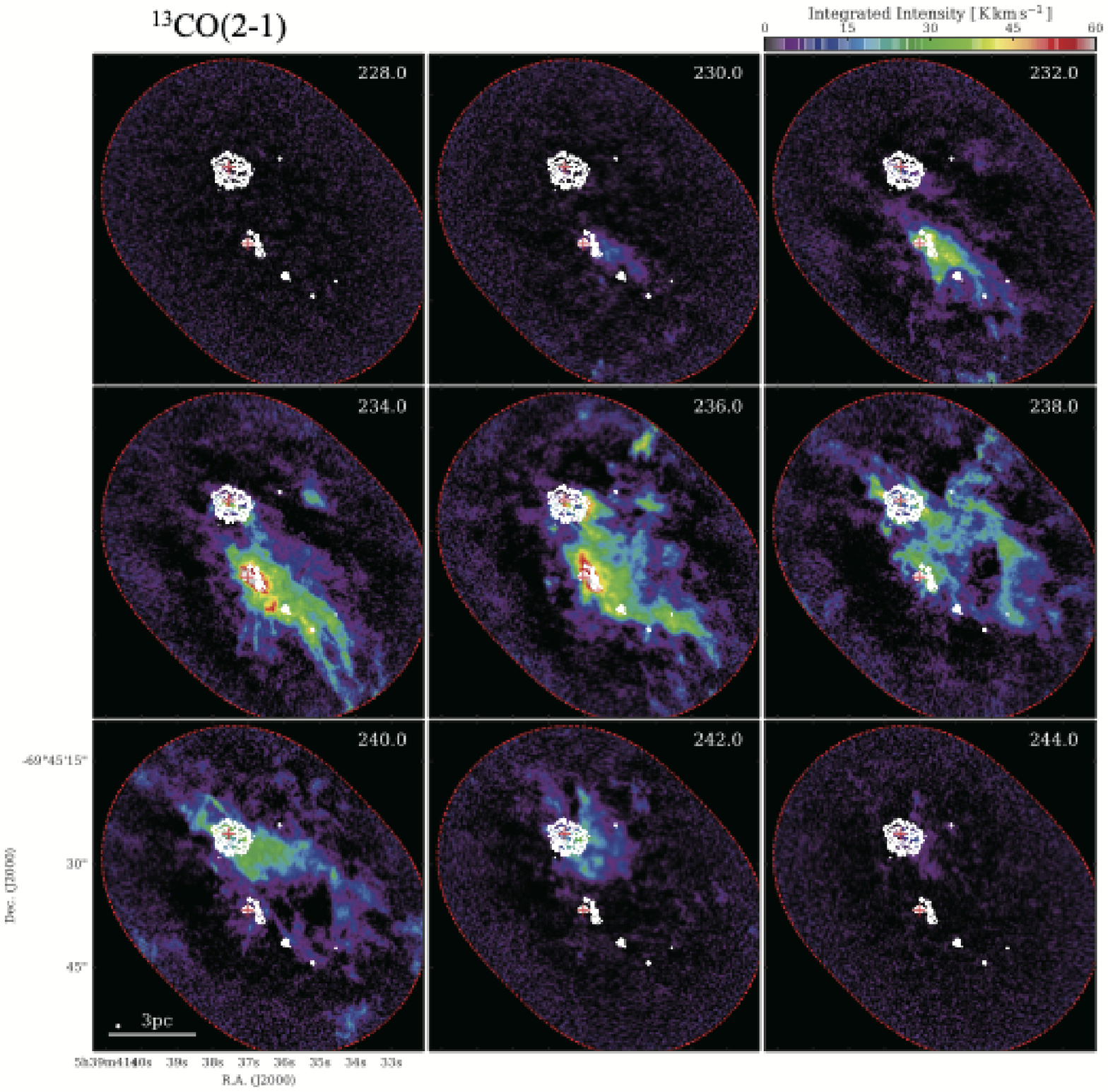}
\caption{Same as Figure~\ref{fig:12COchan}, but for $^{13}$CO($J$ = 2--1).
\label{fig:13COchan}}
\end{figure*}

Figures~\ref{fig:12COchan} and \ref{fig:13COchan} show the $^{12}$CO($J$ = 2--1) and $^{13}$CO($J$ = 2--1) channel maps, respectively, with a velocity bin of 2\,km\,s$^{-1}$. 

\section{Outflow spectra and properties}\label{A:outflow}

\begin{figure*}[htbp]
\centering
\includegraphics[width=180mm]{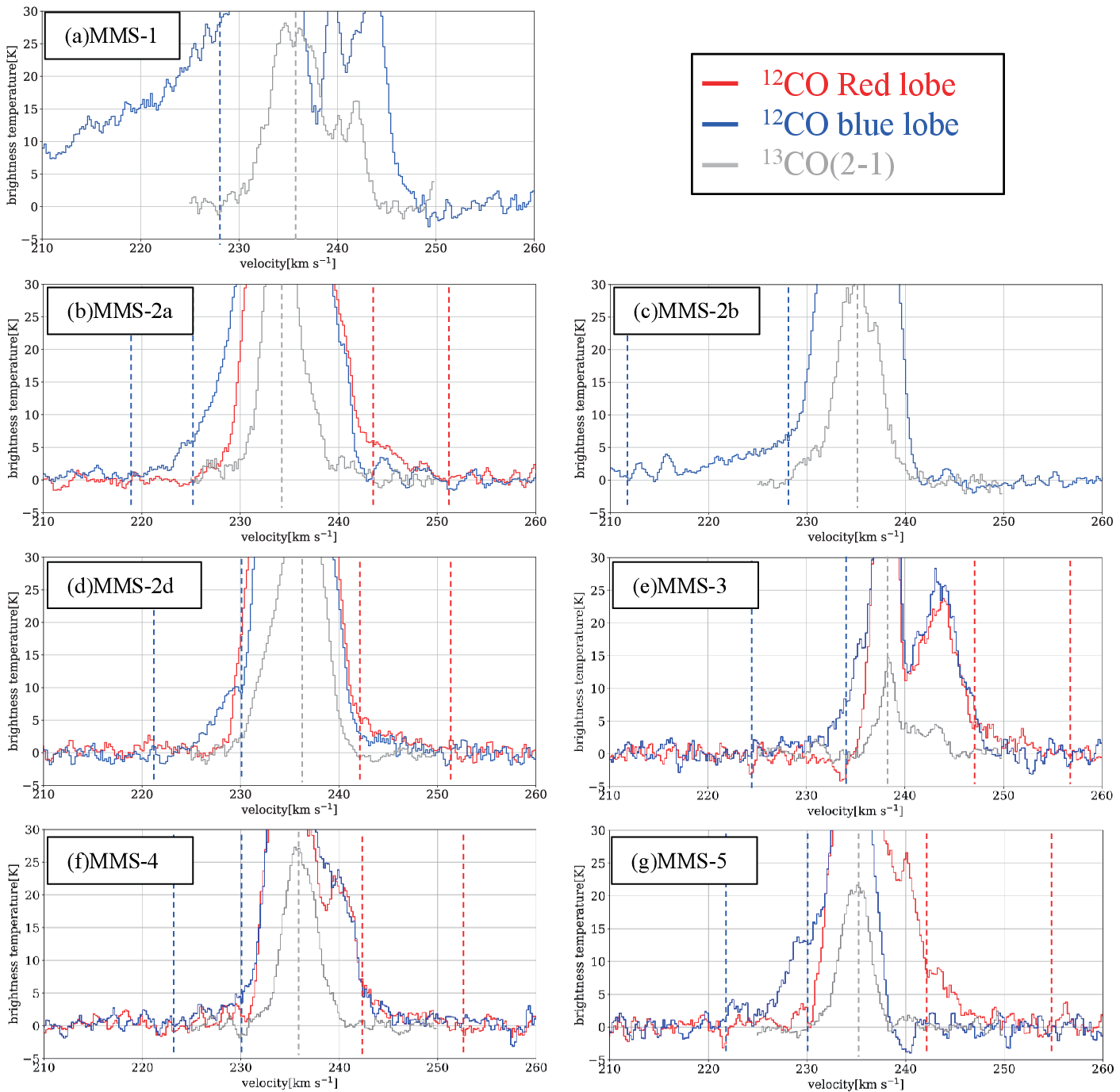}
\caption{$^{12}$CO spectra with high-velocity wing components toward millimeter continuum sources in N159W-North. Red and blue profiles show the $^{12}$CO redshifted and blueshifted components, respectively, averaged within the red and blue contours in Figure~\ref{fig:Cy4cont}b--f. The red and blue vertical lines indicate the velocity ranges of high-velocity emission (see also Table~\ref{tab:MMS_outflow}). 
Gray profiles show $^{13}$CO($J$ = 2--1) spectra at the MMS peaks. Gray dotted vertical lines represent the central velocity determined by Gaussian fitting the $^{13}$CO spectra with a single Gaussian profile.  
%outflow spectra with MMS sources in N159W-North  
\label{fig:outflow_spect}}
\end{figure*}

Figure~\ref{fig:outflow_spect} shows $^{12}$CO spectra with high-velocity wing emission originated from protostellar sources in MMSs. Table~\ref{tab:MMS_outflow} lists the outflow properties.  

\begin{table*}[htbp]
    %\centering
    \begin{flushleft}
     \caption{Outflow properties associated with millimeter sources (MMSs) in N159W-North}
    \begin{tabular}{lccccccc} \hline \hline
    \multicolumn{1}{l}{Source} & \multicolumn{3}{c}{Blue lobe} &  &\multicolumn{3}{c}{Red lobe} \\  \cline{2-4} \cline{6-8}
    Name & $L_{\rm CO}$ [K\,km\,s$^{-1}$\,pc$^{2}$]$^{\rm a}$ & $v_{\rm max}$ [km s$^{-1}$]$^{\rm b}$ & $v_{\rm range}$ [km s$^{-1}$] & & $L_{\rm CO}$ [K\,km\,s$^{-1}$\,pc$^{2}$]$^{\rm a}$ & $v_{\rm max}$ [km s$^{-1}$]$^{\rm b}$ & $v_{\rm range}$ [km s$^{-1}$] \\ \hline
    MMS-1 & 2.87 & $>$24.8 & $<$210--231 & & $\cdots$ & $\cdots$ & $\cdots$ \\
    MMS-2a & 0.57 & 15.8 & 218--225 & & 0.47 & 18.0 & 242--252   \\
    MMS-2b & 0.27 & 18.0 & 217--228 & & $\cdots$ & $\cdots$ & $\cdots$ \\
    MMS-2d & 0.05 & 12.8 & 223--229 & & 0.70 & 14.2 & 242--250   \\
    MMS-3 & 0.21 & 14.6 & 224-234 & & 0.48 & 18.4 & 247-257    \\
    MMS-4 & 0.04 & 12.8 & 223--230 & & 0.27 & 16.2 & 243--252   \\
    MMS-5 & 0.23 & 11.8 & 223--230 & & 0.20 & 21.2 & 243--253    \\
    \hline
    \end{tabular}
    \tablenotetext{\rm a}{Total CO($J$ = 2--1) luminosity of outflow lobes above $\sim$3$\sigma$ detection.}
    \tablenotetext{\rm b}{Maximum radial velocity of outflow lobes with respect to the systemic velocity determined with the $^{13}$CO(2--1) spectra.}
    \label{tab:MMS_outflow}
    \end{flushleft}
\end{table*}

\section{Large-scale CO channel maps}\label{A:Cy7chanmap}

\begin{figure*}[htbp]
\centering
\includegraphics[width=180mm]{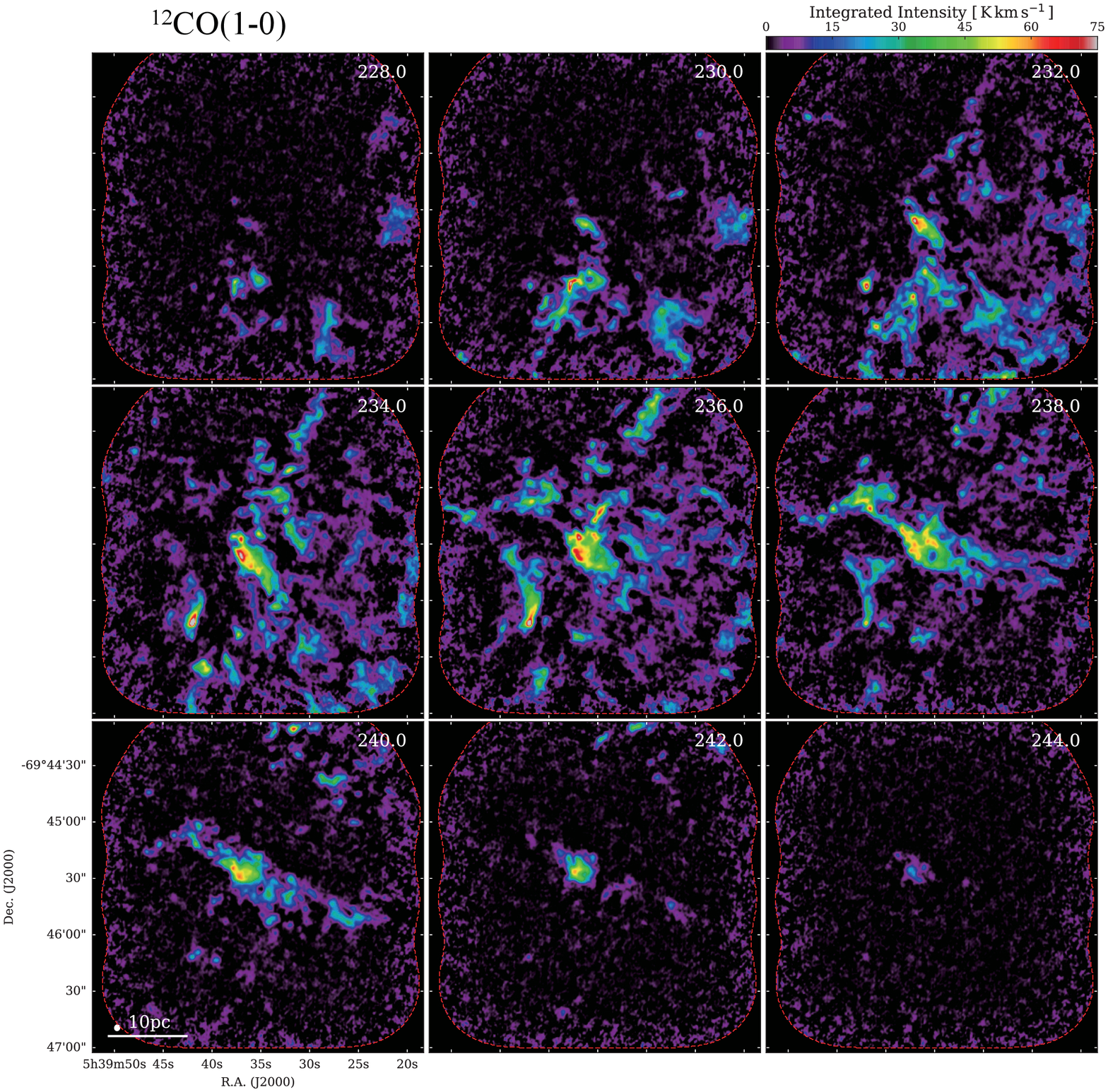}
\caption{Velocity channel maps of N159W in $^{12}$CO($J$ = 1--0). The lowest velocity in unit of km\,s$^{-1}$ of each panel are given in the upper right corners. The white ellipse in the lower left corner in the lower left panel shows the beam size, 2\farcs2 $\times$ 1\farcs8, of the $^{12}$CO data. Red dotted lines show the field coverage.
\label{fig:CO10chan}}
\end{figure*}

Figure~\ref{fig:CO10chan} illustrates the velocity channel maps of $^{12}$CO($J$ = 1--0) across the N159W region.

\section{C$^{18}$O dense clumps traced in the N159E-Papillon and W-South regions}\label{A:C18O}

\begin{figure*}[htbp]
\centering
\includegraphics[width=130mm]{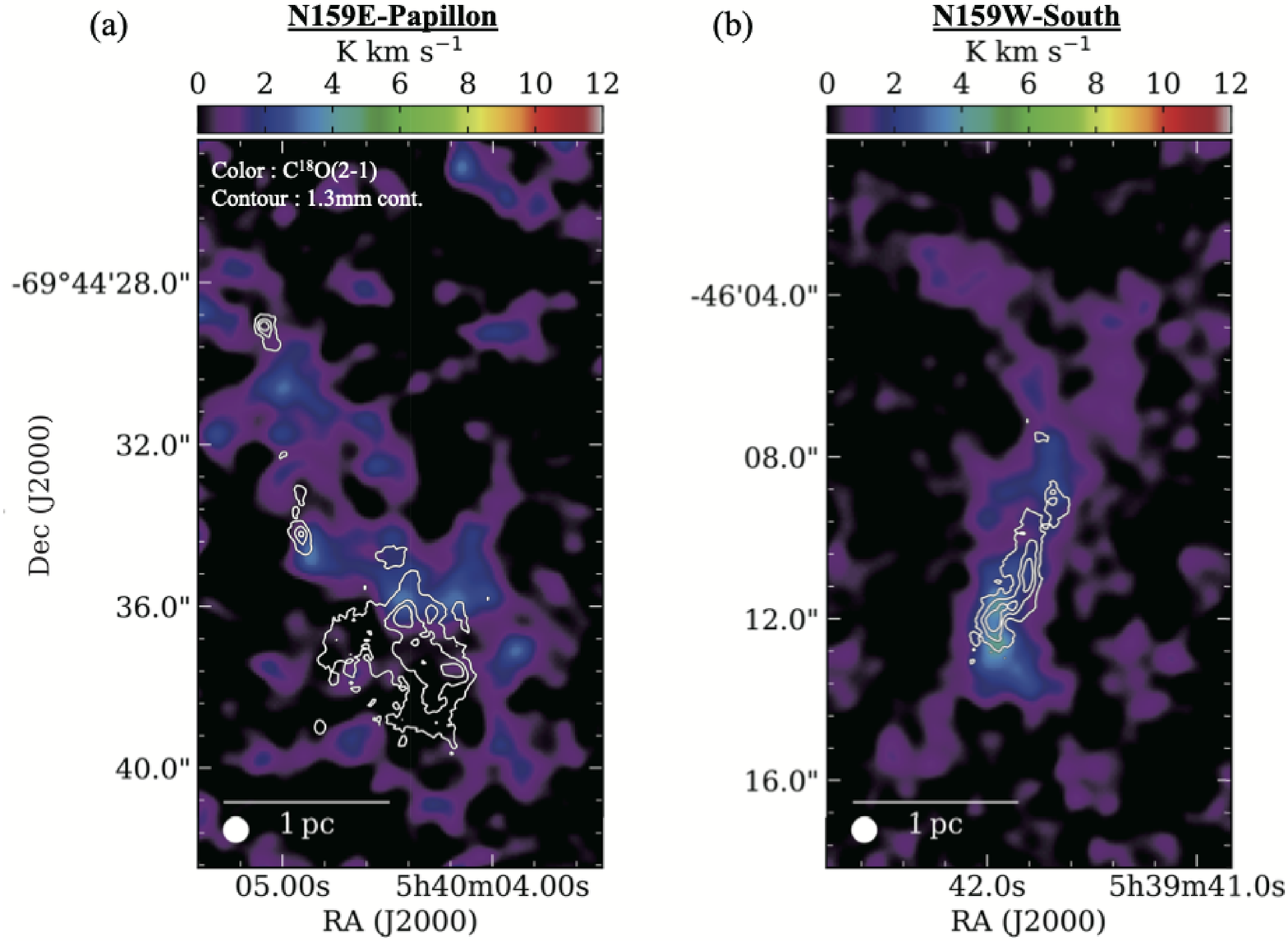}
\caption{C$^{18}$O and 1.3\,mm continuum emission toward N159E-Papillon and N159W-South regions (Papers~I; II). (a) The color-scale image shows the C$^{18}$O velocity-integrated intensity map of N159E-Papillon with an angular resolution of 0\farcs6 shown in the lower left corner. The white contours show the 1.3\,mm continuum emission. The lowest and subsequent contour steps are 0.15\,mJy\,beam$^{-1}$. Note that we coordinate to the intensity scale with that in Figure~\ref{fig:MMS2C18O}a for the comparison purpose. (b) Same as panel (a), but for N159W-South.
\label{fig:C18O}}
\end{figure*}

We compare the C$^{18}$O distribution in N159W-North and that in the two dense clumps in N159E-Papillon (Figure~\ref{fig:C18O}a) and N159W-South (Figure~\ref{fig:C18O}b). Both C$^{18}$O emitting regions are elongated in the direction along the major axis of the filamentary structures traced in $^{12}$CO and $^{13}$CO (Figure~\ref{fig:N159EW_Cy1_4}, see also Papers~I and II). The emitting region reaches a high-density of $\sim$10$^{6}$\,cm$^{-3}$ at maximum in the N159W-South region (Paper~I). However, the peak integrated intensities are $\sim$1--2\,K\, which are significantly lower than those in N159W-North. The currently available ALMA surveys (\citealt{Fukui_2015,Saigo_2017}; Papers~I and II) detected C$^{18}$O emission only in these three spots, N159W-North/South, and E-Papillon, across the N159E/W region, indicating that the C$^{18}$O detection itself is highly rare in the LMC (see also the 30~Dor case in Figure~7 of \citealt{Indebetouw_2013}). Among all known C$^{18}$O emitters, the N159W-North clump is the strongest one in the LMC.

%\label{sec:reference}


\bibliographystyle{aasjournal}
\bibliography{N159W_N_reference.bib}

\begin{thebibliography}{}
\expandafter\ifx\csname natexlab\endcsname\relax\def\natexlab#1{#1}\fi
\providecommand{\url}[1]{\href{#1}{#1}}
\providecommand{\dodoi}[1]{doi:~\href{http://doi.org/#1}{\nolinkurl{#1}}}
\providecommand{\doeprint}[1]{\href{http://ascl.net/#1}{\nolinkurl{http://ascl.net/#1}}}
\providecommand{\doarXiv}[1]{\href{https://arxiv.org/abs/#1}{\nolinkurl{https://arxiv.org/abs/#1}}}

\bibitem[{{Abe} {et~al.}(2021){Abe}, {Inoue}, {Inutsuka}, \&
  {Matsumoto}}]{Abe_2021}
{Abe}, D., {Inoue}, T., {Inutsuka}, S.-i., \& {Matsumoto}, T. 2021, \apj, 916,
  83, \dodoi{10.3847/1538-4357/ac07a1}

\bibitem[{{Astropy Collaboration} {et~al.}(2018){Astropy Collaboration},
  {Price-Whelan}, {Sip{\H{o}}cz}, {G{\"u}nther}, {Lim}, {Crawford}, {Conseil},
  {Shupe}, {Craig}, {Dencheva}, {Ginsburg}, {VanderPlas}, {Bradley},
  {P{\'e}rez-Su{\'a}rez}, {de Val-Borro}, {Aldcroft}, {Cruz}, {Robitaille},
  {Tollerud}, {Ardelean}, {Babej}, {Bach}, {Bachetti}, {Bakanov}, {Bamford},
  {Barentsen}, {Barmby}, {Baumbach}, {Berry}, {Biscani}, {Boquien}, {Bostroem},
  {Bouma}, {Brammer}, {Bray}, {Breytenbach}, {Buddelmeijer}, {Burke},
  {Calderone}, {Cano Rodr{\'\i}guez}, {Cara}, {Cardoso}, {Cheedella}, {Copin},
  {Corrales}, {Crichton}, {D'Avella}, {Deil}, {Depagne}, {Dietrich}, {Donath},
  {Droettboom}, {Earl}, {Erben}, {Fabbro}, {Ferreira}, {Finethy}, {Fox},
  {Garrison}, {Gibbons}, {Goldstein}, {Gommers}, {Greco}, {Greenfield},
  {Groener}, {Grollier}, {Hagen}, {Hirst}, {Homeier}, {Horton}, {Hosseinzadeh},
  {Hu}, {Hunkeler}, {Ivezi{\'c}}, {Jain}, {Jenness}, {Kanarek}, {Kendrew},
  {Kern}, {Kerzendorf}, {Khvalko}, {King}, {Kirkby}, {Kulkarni}, {Kumar},
  {Lee}, {Lenz}, {Littlefair}, {Ma}, {Macleod}, {Mastropietro}, {McCully},
  {Montagnac}, {Morris}, {Mueller}, {Mumford}, {Muna}, {Murphy}, {Nelson},
  {Nguyen}, {Ninan}, {N{\"o}the}, {Ogaz}, {Oh}, {Parejko}, {Parley}, {Pascual},
  {Patil}, {Patil}, {Plunkett}, {Prochaska}, {Rastogi}, {Reddy Janga},
  {Sabater}, {Sakurikar}, {Seifert}, {Sherbert}, {Sherwood-Taylor}, {Shih},
  {Sick}, {Silbiger}, {Singanamalla}, {Singer}, {Sladen}, {Sooley},
  {Sornarajah}, {Streicher}, {Teuben}, {Thomas}, {Tremblay}, {Turner},
  {Terr{\'o}n}, {van Kerkwijk}, {de la Vega}, {Watkins}, {Weaver}, {Whitmore},
  {Woillez}, {Zabalza}, \& {Astropy Contributors}}]{Astropy18}
{Astropy Collaboration}, {Price-Whelan}, A.~M., {Sip{\H{o}}cz}, B.~M., {et~al.}
  2018, \aj, 156, 123, \dodoi{10.3847/1538-3881/aabc4f}

\bibitem[{{Bailer-Jones} {et~al.}(2018){Bailer-Jones}, {Rybizki}, {Fouesneau},
  {Mantelet}, \& {Andrae}}]{Bailer-Jones_2018}
{Bailer-Jones}, C.~A.~L., {Rybizki}, J., {Fouesneau}, M., {Mantelet}, G., \&
  {Andrae}, R. 2018, \aj, 156, 58, \dodoi{10.3847/1538-3881/aacb21}

\bibitem[{Balbinot {et~al.}(2015)Balbinot, Santiago, Girardi, Pieres, da~Costa,
  Maia, Gruendl, Walker, Yanny, Drlica-Wagner, Benoit-Levy, Abbott, Allam,
  Annis, Bernstein, Bernstein, Bertin, Brooks, Buckley-Geer, Rosell, Cunha,
  DePoy, Desai, Diehl, Doel, Estrada, Evrard, Neto, Finley, Flaugher, Frieman,
  Gruen, Honscheid, James, Kuehn, Kuropatkin, Lahav, March, Marshall, Miller,
  Miquel, Ogando, Peoples, Plazas, Scarpine, Schubnell, Sevilla-Noarbe, Smith,
  Soares-Santos, Suchyta, Swanson, Tarle, Tucker, Wechsler, \&
  Zuntz}]{Balbinot_2015}
Balbinot, E., Santiago, B.~X., Girardi, L., {et~al.} 2015, Monthly Notices of
  the Royal Astronomical Society, 449, 1129, \dodoi{10.1093/mnras/stv356}

\bibitem[{{Bernard} {et~al.}(2016){Bernard}, {Neichel}, {Samal}, {Zavagno},
  {Andersen}, {Evans}, {Plana}, \& {Fusco}}]{Bernard_2016}
{Bernard}, A., {Neichel}, B., {Samal}, M.~R., {et~al.} 2016, \aap, 592, A77,
  \dodoi{10.1051/0004-6361/201628754}

\bibitem[{{Beuther} {et~al.}(2002){Beuther}, {Schilke}, {Sridharan}, {Menten},
  {Walmsley}, \& {Wyrowski}}]{Beuther_2002}
{Beuther}, H., {Schilke}, P., {Sridharan}, T.~K., {et~al.} 2002, \aap, 383,
  892, \dodoi{10.1051/0004-6361:20011808}

\bibitem[{{Carlson} {et~al.}(2012){Carlson}, {Sewi{\l}o}, {Meixner}, {Romita},
  \& {Lawton}}]{Carlson_2012}
{Carlson}, L.~R., {Sewi{\l}o}, M., {Meixner}, M., {Romita}, K.~A., \& {Lawton},
  B. 2012, \aap, 542, A66, \dodoi{10.1051/0004-6361/201118627}

\bibitem[{{Ceccarelli} {et~al.}(2007){Ceccarelli}, {Caselli}, {Herbst},
  {Tielens}, \& {Caux}}]{Ceccarelli_2007}
{Ceccarelli}, C., {Caselli}, P., {Herbst}, E., {Tielens}, A.~G.~G.~M., \&
  {Caux}, E. 2007, in Protostars and Planets V, ed. B.~{Reipurth}, D.~{Jewitt},
  \& K.~{Keil}, 47.
\newblock \doarXiv{astro-ph/0603018}

\bibitem[{Chen {et~al.}(2010)Chen, Indebetouw, Chu, Gruendl, Testor, Heitsch,
  Seale, Meixner, \& Sewilo}]{Chen_2010}
Chen, C.-H.~R., Indebetouw, R., Chu, Y.-H., {et~al.} 2010, The Astrophysical
  Journal, 721, 1206, \dodoi{10.1088/0004-637x/721/2/1206}

\bibitem[{{Cignoni} {et~al.}(2015){Cignoni}, {Sabbi}, {van der Marel}, {Tosi},
  {Zaritsky}, {Anderson}, {Lennon}, {Aloisi}, {de Marchi}, {Gouliermis},
  {Grebel}, {Smith}, \& {Zeidler}}]{Cignoni_2015}
{Cignoni}, M., {Sabbi}, E., {van der Marel}, R.~P., {et~al.} 2015, \apj, 811,
  76, \dodoi{10.1088/0004-637X/811/2/76}

\bibitem[{de~Grijs {et~al.}(2014)de~Grijs, Wicker, \& Bono}]{de_Grijs_2014}
de~Grijs, R., Wicker, J.~E., \& Bono, G. 2014, The Astronomical Journal, 147,
  122, \dodoi{10.1088/0004-6256/147/5/122}

\bibitem[{{Dobashi} {et~al.}(2019){Dobashi}, {Shimoikura}, {Katakura},
  {Nakamura}, \& {Shimajiri}}]{Dobashi_2019}
{Dobashi}, K., {Shimoikura}, T., {Katakura}, S., {Nakamura}, F., \&
  {Shimajiri}, Y. 2019, \pasj, 71, S12, \dodoi{10.1093/pasj/psz041}

\bibitem[{{Frerking} {et~al.}(1982){Frerking}, {Langer}, \&
  {Wilson}}]{Frerking_1982}
{Frerking}, M.~A., {Langer}, W.~D., \& {Wilson}, R.~W. 1982, \apj, 262, 590,
  \dodoi{10.1086/160451}

\bibitem[{{Fujita} {et~al.}(2021){Fujita}, {Torii}, {Kuno}, {Nishimura},
  {Umemoto}, {Minamidani}, {Kohno}, {Yamagishi}, {Tosaki}, {Matsuo}, {Tsuda},
  {Enokiya}, {Tachihara}, {Ohama}, {Sano}, {Okawa}, {Hayashi}, {Yoshiike},
  {Tsutsumi}, \& {Fukui}}]{Fujita_2021}
{Fujita}, S., {Torii}, K., {Kuno}, N., {et~al.} 2021, \pasj, 73, S172,
  \dodoi{10.1093/pasj/psz028}

\bibitem[{{Fukui} {et~al.}(2021){Fukui}, {Habe}, {Inoue}, {Enokiya}, \&
  {Tachihara}}]{Fukui_2021}
{Fukui}, Y., {Habe}, A., {Inoue}, T., {Enokiya}, R., \& {Tachihara}, K. 2021,
  \pasj, 73, S1, \dodoi{10.1093/pasj/psaa103}

\bibitem[{Fukui {et~al.}(2017)Fukui, Tsuge, Sano, Bekki, Yozin, Tachihara, \&
  Inoue}]{Fukui_2017}
Fukui, Y., Tsuge, K., Sano, H., {et~al.} 2017, Publications of the Astronomical
  Society of Japan, 69, \dodoi{10.1093/pasj/psx032}

\bibitem[{Fukui {et~al.}(2008)Fukui, Kawamura, Minamidani, Mizuno, Kanai,
  Mizuno, Onishi, Yonekura, Mizuno, Ogawa, \& Rubio}]{Fukui_2008}
Fukui, Y., Kawamura, A., Minamidani, T., {et~al.} 2008, The Astrophysical
  Journal Supplement Series, 178, 56, \dodoi{10.1086/589833}

\bibitem[{Fukui {et~al.}(2015)Fukui, Harada, Tokuda, Morioka, Onishi, Torii,
  Ohama, Hattori, Nayak, Meixner, Sewi{\l}o, Indebetouw, Kawamura, Saigo,
  Yamamoto, Tachihara, Minamidani, Inoue, Madden, Galametz, Lebouteiller,
  Mizuno, \& Chen}]{Fukui_2015}
Fukui, Y., Harada, R., Tokuda, K., {et~al.} 2015, The Astrophysical Journal,
  807, L4, \dodoi{10.1088/2041-8205/807/1/l4}

\bibitem[{Fukui {et~al.}(2018)Fukui, Torii, Hattori, Nishimura, Ohama,
  Shimajiri, Shima, Habe, Sano, Kohno, Yamamoto, Tachihara, \&
  Onishi}]{Fukui_2018a}
Fukui, Y., Torii, K., Hattori, Y., {et~al.} 2018, The Astrophysical Journal,
  859, 166, \dodoi{10.3847/1538-4357/aac217}

\bibitem[{Fukui {et~al.}(2019)Fukui, Tokuda, Saigo, Harada, Tachihara, Tsuge,
  Inoue, Torii, Nishimura, Zahorecz, Nayak, Meixner, Minamidani, Kawamura,
  Mizuno, Indebetouw, Sewi{\l}o, Madden, Galametz, Lebouteiller, Chen, \&
  Onishi}]{Fukui_2019}
Fukui, Y., Tokuda, K., Saigo, K., {et~al.} 2019, The Astrophysical Journal,
  886, 14, \dodoi{10.3847/1538-4357/ab4900}

\bibitem[{{Fukushima} \& {Yajima}(2021)}]{Fukushima_2021}
{Fukushima}, H., \& {Yajima}, H. 2021, \mnras, 506, 5512,
  \dodoi{10.1093/mnras/stab2099}

\bibitem[{{Galametz} {et~al.}(2020){Galametz}, {Schruba}, {De Breuck}, {Immer},
  {Chevance}, {Galliano}, {Gusdorf}, {Lebouteiller}, {Lee}, {Madden}, {Polles},
  \& {van Kempen}}]{Galametz_2020}
{Galametz}, M., {Schruba}, A., {De Breuck}, C., {et~al.} 2020, \aap, 643, A63,
  \dodoi{10.1051/0004-6361/202038641}

\bibitem[{{Galv{\'a}n-Madrid} {et~al.}(2013){Galv{\'a}n-Madrid}, {Liu},
  {Zhang}, {Pineda}, {Peng}, {Zhang}, {Keto}, {Ho}, {Rodr{\'\i}guez}, {Zapata},
  {Peters}, \& {De Pree}}]{2013ApJ...779..121G}
{Galv{\'a}n-Madrid}, R., {Liu}, H.~B., {Zhang}, Z.~Y., {et~al.} 2013, \apj,
  779, 121, \dodoi{10.1088/0004-637X/779/2/121}

\bibitem[{{Ginsburg} {et~al.}(2012){Ginsburg}, {Bressert}, {Bally}, \&
  {Battersby}}]{Ginsburg_2012}
{Ginsburg}, A., {Bressert}, E., {Bally}, J., \& {Battersby}, C. 2012, \apjl,
  758, L29, \dodoi{10.1088/2041-8205/758/2/L29}

\bibitem[{Gordon {et~al.}(2014)Gordon, Roman-Duval, Bot, Meixner, Babler,
  Bernard, Bolatto, Boyer, Clayton, Engelbracht, Fukui, Galametz, Galliano,
  Hony, Hughes, Indebetouw, Israel, Jameson, Kawamura, Lebouteiller, Li,
  Madden, Matsuura, Misselt, Montiel, Okumura, Onishi, Panuzzo, Paradis, Rubio,
  Sandstrom, Sauvage, Seale, Sewi{\l}o, Tchernyshyov, \& Skibba}]{Gordon_2014}
Gordon, K.~D., Roman-Duval, J., Bot, C., {et~al.} 2014, The Astrophysical
  Journal, 797, 85, \dodoi{10.1088/0004-637x/797/2/85}

\bibitem[{{Gruendl} \& {Chu}(2009)}]{Gruendl_2009}
{Gruendl}, R.~A., \& {Chu}, Y.-H. 2009, \apjs, 184, 172,
  \dodoi{10.1088/0067-0049/184/1/172}

\bibitem[{Herrera {et~al.}(2013)Herrera, Rubio, Bolatto, Boulanger, Israel, \&
  Rantakyr{\"o}}]{Herrera_2013}
Herrera, C.~N., Rubio, M., Bolatto, A.~D., {et~al.} 2013, Astronomy {\&}
  Astrophysics, 554, A91, \dodoi{10.1051/0004-6361/201219381}

\bibitem[{{Hirota} {et~al.}(2017){Hirota}, {Machida}, {Matsushita}, {Motogi},
  {Matsumoto}, {Kim}, {Burns}, \& {Honma}}]{Hirota_2017}
{Hirota}, T., {Machida}, M.~N., {Matsushita}, Y., {et~al.} 2017, Nature
  Astronomy, 1, 0146, \dodoi{10.1038/s41550-017-0146}

\bibitem[{{Hodge}(1961)}]{Hodge_1961}
{Hodge}, P.~W. 1961, \apj, 133, 413, \dodoi{10.1086/147044}

\bibitem[{{Hughes} {et~al.}(2010){Hughes}, {Wong}, {Ott}, {Muller}, {Pineda},
  {Mizuno}, {Bernard}, {Paradis}, {Maddison}, {Reach}, {Staveley-Smith},
  {Kawamura}, {Meixner}, {Kim}, {Onishi}, {Mizuno}, \& {Fukui}}]{Hughes_2010}
{Hughes}, A., {Wong}, T., {Ott}, J., {et~al.} 2010, \mnras, 406, 2065,
  \dodoi{10.1111/j.1365-2966.2010.16829.x}

\bibitem[{{Hunter} {et~al.}(2003){Hunter}, {Elmegreen}, {Dupuy}, \&
  {Mortonson}}]{Hunter03}
{Hunter}, D.~A., {Elmegreen}, B.~G., {Dupuy}, T.~J., \& {Mortonson}, M. 2003,
  \aj, 126, 1836, \dodoi{10.1086/378056}

\bibitem[{{Indebetouw} {et~al.}(2004){Indebetouw}, {Johnson}, \&
  {Conti}}]{Indebetouw_2004}
{Indebetouw}, R., {Johnson}, K.~E., \& {Conti}, P. 2004, \aj, 128, 2206,
  \dodoi{10.1086/424614}

\bibitem[{{Indebetouw} {et~al.}(2020){Indebetouw}, {Wong}, {Chen}, {Kepley},
  {Lebouteiller}, {Madden}, \& {Oliveira}}]{Indebetouw_2020}
{Indebetouw}, R., {Wong}, T., {Chen}, C. H.~R., {et~al.} 2020, \apj, 888, 56,
  \dodoi{10.3847/1538-4357/ab5db7}

\bibitem[{Indebetouw {et~al.}(2013)Indebetouw, Brogan, Chen, Leroy, Johnson,
  Muller, Madden, Cormier, Galliano, Hughes, Hunter, Kawamura, Kepley,
  Lebouteiller, Meixner, Oliveira, Onishi, \& Vasyunina}]{Indebetouw_2013}
Indebetouw, R., Brogan, C., Chen, C.-H.~R., {et~al.} 2013, The Astrophysical
  Journal, 774, 73, \dodoi{10.1088/0004-637x/774/1/73}

\bibitem[{{Inoue} {et~al.}(2018){Inoue}, {Hennebelle}, {Fukui}, {Matsumoto},
  {Iwasaki}, \& {Inutsuka}}]{Inoue_2018}
{Inoue}, T., {Hennebelle}, P., {Fukui}, Y., {et~al.} 2018, \pasj, 70, S53,
  \dodoi{10.1093/pasj/psx089}

\bibitem[{{Johansson} {et~al.}(1994){Johansson}, {Olofsson}, {Hjalmarson},
  {Gredel}, \& {Black}}]{Johansson_1994}
{Johansson}, L.~E.~B., {Olofsson}, H., {Hjalmarson}, A., {Gredel}, R., \&
  {Black}, J.~H. 1994, \aap, 291, 89

\bibitem[{{Johansson} {et~al.}(1998){Johansson}, {Greve}, {Booth}, {Boulanger},
  {Garay}, {de Graauw}, {Israel}, {Kutner}, {Lequeux}, {Murphy}, {Nyman}, \&
  {Rubio}}]{1998A&A...331..857J}
{Johansson}, L.~E.~B., {Greve}, A., {Booth}, R.~S., {et~al.} 1998, \aap, 331,
  857

\bibitem[{{Jones} {et~al.}(2017){Jones}, {Woods}, {Kemper}, {Kraemer}, {Sloan},
  {Srinivasan}, {Oliveira}, {van Loon}, {Boyer}, {Sargent}, {McDonald},
  {Meixner}, {Zijlstra}, {Ruffle}, {Lagadec}, {Pauly}, {Sewi{\l}o}, {Clayton},
  \& {Volk}}]{Jones_2017}
{Jones}, O.~C., {Woods}, P.~M., {Kemper}, F., {et~al.} 2017, \mnras, 470, 3250,
  \dodoi{10.1093/mnras/stx1101}

\bibitem[{Kawamura {et~al.}(2009)Kawamura, Mizuno, Minamidani,
  Fillipovi{\'{c}}, Staveley-Smith, Kim, Mizuno, Onishi, Mizuno, \&
  Fukui}]{Kawamura_2009}
Kawamura, A., Mizuno, Y., Minamidani, T., {et~al.} 2009, The Astrophysical
  Journal Supplement Series, 184, 1, \dodoi{10.1088/0067-0049/184/1/1}

\bibitem[{{Kohno} {et~al.}(2021){Kohno}, {Tachihara}, {Torii}, {Fujita},
  {Nishimura}, {Kuno}, {Umemoto}, {Minamidani}, {Matsuo}, {Kiridoshi},
  {Tokuda}, {Hanaoka}, {Tsuda}, {Kuriki}, {Ohama}, {Sano}, {Hasegawa}, {Sofue},
  {Habe}, {Onishi}, \& {Fukui}}]{Kohno_2021}
{Kohno}, M., {Tachihara}, K., {Torii}, K., {et~al.} 2021, \pasj, 73, S129,
  \dodoi{10.1093/pasj/psaa015}

\bibitem[{{Kong} {et~al.}(2019){Kong}, {Arce}, {Maureira}, {Caselli}, {Tan}, \&
  {Fontani}}]{Kong_2019}
{Kong}, S., {Arce}, H.~G., {Maureira}, M.~J., {et~al.} 2019, \apj, 874, 104,
  \dodoi{10.3847/1538-4357/ab07b9}

\bibitem[{{Kong} {et~al.}(2017){Kong}, {Tan}, {Caselli}, {Fontani}, {Liu}, \&
  {Butler}}]{Kong_2017}
{Kong}, S., {Tan}, J.~C., {Caselli}, P., {et~al.} 2017, \apj, 834, 193,
  \dodoi{10.3847/1538-4357/834/2/193}

\bibitem[{{Krumholz} \& {McKee}(2020)}]{Krumholz_2020}
{Krumholz}, M.~R., \& {McKee}, C.~F. 2020, \mnras, 494, 624,
  \dodoi{10.1093/mnras/staa659}

\bibitem[{{Kumai} {et~al.}(1993){Kumai}, {Basu}, \& {Fujimoto}}]{Kumai93}
{Kumai}, Y., {Basu}, B., \& {Fujimoto}, M. 1993, \apj, 404, 144,
  \dodoi{10.1086/172265}

\bibitem[{{Lada} \& {Lada}(2003)}]{Lada_2003}
{Lada}, C.~J., \& {Lada}, E.~A. 2003, \araa, 41, 57,
  \dodoi{10.1146/annurev.astro.41.011802.094844}

\bibitem[{Lee {et~al.}(2012)Lee, Murray, \& Rahman}]{Lee_2012}
Lee, E.~J., Murray, N., \& Rahman, M. 2012, The Astrophysical Journal, 752,
  146, \dodoi{10.1088/0004-637x/752/2/146}

\bibitem[{{Lee} {et~al.}(2016){Lee}, {Madden}, {Lebouteiller}, {Gusdorf},
  {Godard}, {Wu}, {Galametz}, {Cormier}, {Le Petit}, {Roueff}, {Bron},
  {Carlson}, {Chevance}, {Fukui}, {Galliano}, {Hony}, {Hughes}, {Indebetouw},
  {Israel}, {Kawamura}, {Le Bourlot}, {Lesaffre}, {Meixner}, {Muller}, {Nayak},
  {Onishi}, {Roman-Duval}, \& {Sewi{\l}o}}]{Lee_2016}
{Lee}, M.~Y., {Madden}, S.~C., {Lebouteiller}, V., {et~al.} 2016, \aap, 596,
  A85, \dodoi{10.1051/0004-6361/201628098}

\bibitem[{{Longmore} {et~al.}(2014){Longmore}, {Kruijssen}, {Bastian}, {Bally},
  {Rathborne}, {Testi}, {Stolte}, {Dale}, {Bressert}, \&
  {Alves}}]{Longmore_2014}
{Longmore}, S.~N., {Kruijssen}, J.~M.~D., {Bastian}, N., {et~al.} 2014, in
  Protostars and Planets VI, ed. H.~{Beuther}, R.~S. {Klessen}, C.~P.
  {Dullemond}, \& T.~{Henning}, 291,
  \dodoi{10.2458/azu\_uapress\_9780816531240-ch013}

\bibitem[{Louvet {et~al.}(2014)Louvet, Motte, Hennebelle, Maury, Bonnell,
  Bontemps, Gusdorf, Hill, Gueth, Peretto, Duarte-Cabral, Stephan, Schilke,
  Csengeri, Luong, \& Lis}]{Louvet_2014}
Louvet, F., Motte, F., Hennebelle, P., {et~al.} 2014, Astronomy {\&}
  Astrophysics, 570, A15, \dodoi{10.1051/0004-6361/201423603}

\bibitem[{{Maeda} {et~al.}(2021){Maeda}, {Inoue}, \& {Fukui}}]{Maeda_2021}
{Maeda}, R., {Inoue}, T., \& {Fukui}, Y. 2021, \apj, 908, 2,
  \dodoi{10.3847/1538-4357/abcc75}

\bibitem[{{Matsushita} {et~al.}(2017){Matsushita}, {Machida}, {Sakurai}, \&
  {Hosokawa}}]{Matsushita_2017}
{Matsushita}, Y., {Machida}, M.~N., {Sakurai}, Y., \& {Hosokawa}, T. 2017,
  \mnras, 470, 1026, \dodoi{10.1093/mnras/stx893}

\bibitem[{{Matsushita} {et~al.}(2019){Matsushita}, {Takahashi}, {Machida}, \&
  {Tomisaka}}]{Matsushita_2019}
{Matsushita}, Y., {Takahashi}, S., {Machida}, M.~N., \& {Tomisaka}, K. 2019,
  \apj, 871, 221, \dodoi{10.3847/1538-4357/aaf1b6}

\bibitem[{{McKee} \& {Tan}(2003)}]{McKee_2003}
{McKee}, C.~F., \& {Tan}, J.~C. 2003, \apj, 585, 850, \dodoi{10.1086/346149}

\bibitem[{{McMullin} {et~al.}(2007){McMullin}, {Waters}, {Schiebel}, {Young},
  \& {Golap}}]{McMullin07}
{McMullin}, J.~P., {Waters}, B., {Schiebel}, D., {Young}, W., \& {Golap}, K.
  2007, in Astronomical Society of the Pacific Conference Series, Vol. 376,
  Astronomical Data Analysis Software and Systems XVI, ed. R.~A. {Shaw},
  F.~{Hill}, \& D.~J. {Bell}, 127

\bibitem[{{Meixner} {et~al.}(2006){Meixner}, {Gordon}, {Indebetouw}, {Hora},
  {Whitney}, {Blum}, {Reach}, {Bernard}, {Meade}, {Babler}, {Engelbracht},
  {For}, {Misselt}, {Vijh}, {Leitherer}, {Cohen}, {Churchwell}, {Boulanger},
  {Frogel}, {Fukui}, {Gallagher}, {Gorjian}, {Harris}, {Kelly}, {Kawamura},
  {Kim}, {Latter}, {Madden}, {Markwick-Kemper}, {Mizuno}, {Mizuno}, {Mould},
  {Nota}, {Oey}, {Olsen}, {Onishi}, {Paladini}, {Panagia}, {Perez-Gonzalez},
  {Shibai}, {Sato}, {Smith}, {Staveley-Smith}, {Tielens}, {Ueta}, {van Dyk},
  {Volk}, {Werner}, \& {Zaritsky}}]{Margaret_2006}
{Meixner}, M., {Gordon}, K.~D., {Indebetouw}, R., {et~al.} 2006, \aj, 132,
  2268, \dodoi{10.1086/508185}

\bibitem[{{Millar} \& {Herbst}(1990)}]{Millar_1990}
{Millar}, T.~J., \& {Herbst}, E. 1990, \mnras, 242, 92,
  \dodoi{10.1093/mnras/242.2.92}

\bibitem[{Minamidani {et~al.}(2008)Minamidani, Mizuno, Mizuno, Kawamura,
  Onishi, Hasegawa, Tatematsu, Ikeda, Moriguchi, Yamaguchi, Ott, Wong, Muller,
  Pineda, Hughes, Staveley-Smith, Klein, Mizuno, Nikoli{\'{c}}, Booth,
  Heikkil{\"a}, Nyman, Lerner, Garay, Kim, Fujishita, Kawase, Rubio, \&
  Fukui}]{Minamidani_2008}
Minamidani, T., Mizuno, N., Mizuno, Y., {et~al.} 2008, The Astrophysical
  Journal Supplement Series, 175, 485, \dodoi{10.1086/524038}

\bibitem[{Minamidani {et~al.}(2011)Minamidani, Tanaka, Mizuno, Mizuno,
  Kawamura, Onishi, Hasegawa, Tatematsu, Takekoshi, Sorai, Moribe, Torii,
  Sakai, Muraoka, Tanaka, Ezawa, Kohno, Kim, Rubio, \& Fukui}]{Minamidani_2011}
Minamidani, T., Tanaka, T., Mizuno, Y., {et~al.} 2011, The Astronomical
  Journal, 141, 73, \dodoi{10.1088/0004-6256/141/3/73}

\bibitem[{{Miyawaki} {et~al.}(2009){Miyawaki}, {Hayashi}, \&
  {Hasegawa}}]{Miyawaki_2009}
{Miyawaki}, R., {Hayashi}, M., \& {Hasegawa}, T. 2009, \pasj, 61, 39,
  \dodoi{10.1093/pasj/61.1.39}

\bibitem[{{Miyawaki} {et~al.}(2021){Miyawaki}, {Hayashi}, \&
  {Hasegawa}}]{Miyawaki_2021}
---. 2021, \pasj, \dodoi{10.1093/pasj/psab113}

\bibitem[{{Mizuno} {et~al.}(2010){Mizuno}, {Kawamura}, {Onishi}, {Minamidani},
  {Muller}, {Yamamoto}, {Hayakawa}, {Mizuno}, {Mizuno}, {Stutzki}, {Pineda},
  {Klein}, {Bertoldi}, {Koo}, {Rubio}, {Burton}, {Benz}, {Ezawa}, {Yamaguchi},
  {Kohno}, {Hasegawa}, {Tatematsu}, {Ikeda}, {Ott}, {Wong}, {Hughes},
  {Meixner}, {Indebetouw}, {Gordon}, {Whitney}, {Bernard}, \&
  {Fukui}}]{Mizuno__2010}
{Mizuno}, Y., {Kawamura}, A., {Onishi}, T., {et~al.} 2010, \pasj, 62, 51,
  \dodoi{10.1093/pasj/62.1.51}

\bibitem[{{Molet} {et~al.}(2019){Molet}, {Brouillet}, {Nony}, {Gusdorf},
  {Motte}, {Despois}, {Louvet}, {Bontemps}, \& {Herpin}}]{Molet_2019}
{Molet}, J., {Brouillet}, N., {Nony}, T., {et~al.} 2019, \aap, 626, A132,
  \dodoi{10.1051/0004-6361/201935497}

\bibitem[{{Motogi} {et~al.}(2019){Motogi}, {Hirota}, {Machida}, {Yonekura},
  {Honma}, {Takakuwa}, \& {Matsushita}}]{Motogi_2019}
{Motogi}, K., {Hirota}, T., {Machida}, M.~N., {et~al.} 2019, \apjl, 877, L25,
  \dodoi{10.3847/2041-8213/ab212f}

\bibitem[{Motte {et~al.}(2018)Motte, Bontemps, \& Louvet}]{Motte_2018}
Motte, F., Bontemps, S., \& Louvet, F. 2018, Annual Review of Astronomy and
  Astrophysics, 56, 41, \dodoi{10.1146/annurev-astro-091916-055235}

\bibitem[{Motte {et~al.}(2003)Motte, Schilke, \& Lis}]{Motte_2003}
Motte, F., Schilke, P., \& Lis, D.~C. 2003, The Astrophysical Journal, 582,
  277, \dodoi{10.1086/344538}

\bibitem[{Muraoka {et~al.}(2020)Muraoka, Kondo, Tokuda, Nishimura, Miura,
  Onodera, Kuno, Zahorecz, Tsuge, Sano, Fujita, Onishi, Saigo, Tachihara,
  Fukui, \& Kawamura}]{Muraoka_2020}
Muraoka, K., Kondo, H., Tokuda, K., {et~al.} 2020, The Astrophysical Journal,
  903, 94, \dodoi{10.3847/1538-4357/abb822}

\bibitem[{Myers(2009)}]{Myers_2009}
Myers, P.~C. 2009, The Astrophysical Journal, 700, 1609,
  \dodoi{10.1088/0004-637x/700/2/1609}

\bibitem[{Nayak {et~al.}(2018)Nayak, Meixner, Fukui, Tachihara, Onishi, Saigo,
  Tokuda, \& Harada}]{Nayak_2018}
Nayak, O., Meixner, M., Fukui, Y., {et~al.} 2018, The Astrophysical Journal,
  854, 154, \dodoi{10.3847/1538-4357/aaab5f}

\bibitem[{{Nguyen-Luong} {et~al.}(2016){Nguyen-Luong}, {Nguyen}, {Motte},
  {Schneider}, {Fujii}, {Louvet}, {Hill}, {Sanhueza}, {Chibueze}, \&
  {Didelon}}]{Nguyen_2016}
{Nguyen-Luong}, Q., {Nguyen}, H. V.~V., {Motte}, F., {et~al.} 2016, \apj, 833,
  23, \dodoi{10.3847/0004-637X/833/1/23}

\bibitem[{{Nishimura} {et~al.}(2015){Nishimura}, {Tokuda}, {Kimura}, {Muraoka},
  {Maezawa}, {Ogawa}, {Dobashi}, {Shimoikura}, {Mizuno}, {Fukui}, \&
  {Onishi}}]{Nishimura__2015}
{Nishimura}, A., {Tokuda}, K., {Kimura}, K., {et~al.} 2015, \apjs, 216, 18,
  \dodoi{10.1088/0067-0049/216/1/18}

\bibitem[{{Nishimura} {et~al.}(2016){Nishimura}, {Shimonishi}, {Watanabe},
  {Sakai}, {Aikawa}, {Kawamura}, \& {Yamamoto}}]{Nishimura_2016}
{Nishimura}, Y., {Shimonishi}, T., {Watanabe}, Y., {et~al.} 2016, \apj, 818,
  161, \dodoi{10.3847/0004-637X/818/2/161}

\bibitem[{{Ossenkopf} \& {Henning}(1994)}]{Ossenkopf_1994}
{Ossenkopf}, V., \& {Henning}, T. 1994, \aap, 291, 943

\bibitem[{{Ott} {et~al.}(2010){Ott}, {Henkel}, {Staveley-Smith}, \&
  {Wei{\ss}}}]{Ott2010}
{Ott}, J., {Henkel}, C., {Staveley-Smith}, L., \& {Wei{\ss}}, A. 2010, \apj,
  710, 105, \dodoi{10.1088/0004-637X/710/1/105}

\bibitem[{{Paron} {et~al.}(2016){Paron}, {Ortega}, {Fari{\~n}a}, {Cunningham},
  {Jones}, \& {Rubio}}]{Paron_2016}
{Paron}, S., {Ortega}, M.~E., {Fari{\~n}a}, C., {et~al.} 2016, \mnras, 455,
  518, \dodoi{10.1093/mnras/stv2326}

\bibitem[{Peretto {et~al.}(2013)Peretto, Fuller, Duarte-Cabral, Avison,
  Hennebelle, Pineda, Andr{\'{e}}, Bontemps, Motte, Schneider, \&
  Molinari}]{Peretto_2013}
Peretto, N., Fuller, G.~A., Duarte-Cabral, A., {et~al.} 2013, Astronomy {\&}
  Astrophysics, 555, A112, \dodoi{10.1051/0004-6361/201321318}

\bibitem[{{Portegies Zwart} {et~al.}(2010){Portegies Zwart}, {McMillan}, \&
  {Gieles}}]{Portegies_2010}
{Portegies Zwart}, S.~F., {McMillan}, S. L.~W., \& {Gieles}, M. 2010, \araa,
  48, 431, \dodoi{10.1146/annurev-astro-081309-130834}

\bibitem[{{Robitaille} \& {Bressert}(2012)}]{Robi12}
{Robitaille}, T., \& {Bressert}, E. 2012, {APLpy: Astronomical Plotting Library
  in Python}.
\newblock \doeprint{1208.017}

\bibitem[{{Rohlfs} \& {Wilson}(2004)}]{Rohlfs_2004}
{Rohlfs}, K., \& {Wilson}, T.~L. 2004, {Tools of radio astronomy}

\bibitem[{Saigo {et~al.}(2017)Saigo, Onishi, Nayak, Meixner, Tokuda, Harada,
  Morioka, Sewi{\l}o, Indebetouw, Torii, Kawamura, Ohama, Hattori, Yamamoto,
  Tachihara, Minamidani, Inoue, Madden, Galametz, Lebouteiller, Chen, Mizuno,
  \& Fukui}]{Saigo_2017}
Saigo, K., Onishi, T., Nayak, O., {et~al.} 2017, The Astrophysical Journal,
  835, 108, \dodoi{10.3847/1538-4357/835/1/108}

\bibitem[{{Schaefer}(2008)}]{Schaefer_2008}
{Schaefer}, B.~E. 2008, \aj, 135, 112, \dodoi{10.1088/0004-6256/135/1/112}

\bibitem[{Schneider {et~al.}(2010)Schneider, Csengeri, Bontemps, Motte, Simon,
  Hennebelle, Federrath, \& Klessen}]{Schneider_2010}
Schneider, N., Csengeri, T., Bontemps, S., {et~al.} 2010, Astronomy and
  Astrophysics, 520, A49, \dodoi{10.1051/0004-6361/201014481}

\bibitem[{{Seale} {et~al.}(2009){Seale}, {Looney}, {Chu}, {Gruendl}, {Brandl},
  {Chen}, {Brandner}, \& {Blake}}]{Seale_2009}
{Seale}, J.~P., {Looney}, L.~W., {Chu}, Y.-H., {et~al.} 2009, \apj, 699, 150,
  \dodoi{10.1088/0004-637X/699/1/150}

\bibitem[{{Seale} {et~al.}(2014){Seale}, {Meixner}, {Sewi{\l}o}, {Babler},
  {Engelbracht}, {Gordon}, {Hony}, {Misselt}, {Montiel}, {Okumura}, {Panuzzo},
  {Roman-Duval}, {Sauvage}, {Boyer}, {Chen}, {Indebetouw}, {Matsuura},
  {Oliveira}, {Srinivasan}, {van Loon}, {Whitney}, \& {Woods}}]{Seale_2014}
{Seale}, J.~P., {Meixner}, M., {Sewi{\l}o}, M., {et~al.} 2014, \aj, 148, 124,
  \dodoi{10.1088/0004-6256/148/6/124}

\bibitem[{{Sewi{\l}o} {et~al.}(2018){Sewi{\l}o}, {Indebetouw}, {Charnley},
  {Zahorecz}, {Oliveira}, {van Loon}, {Ward}, {Chen}, {Wiseman}, {Fukui},
  {Kawamura}, {Meixner}, {Onishi}, \& {Schilke}}]{2018ApJ...853L..19S}
{Sewi{\l}o}, M., {Indebetouw}, R., {Charnley}, S.~B., {et~al.} 2018, \apjl,
  853, L19, \dodoi{10.3847/2041-8213/aaa079}

\bibitem[{{Sharda} {et~al.}(2021){Sharda}, {Menon}, {Federrath}, {Krumholz},
  {Beattie}, {Jameson}, {Tokuda}, {Burkhart}, {Crocker}, {Law}, {Seta},
  {Gaetz}, {Pingel}, {Seitenzahl}, {Sano}, \& {Fukui}}]{Piyush_2021}
{Sharda}, P., {Menon}, S.~H., {Federrath}, C., {et~al.} 2021, \mnras,
  \dodoi{10.1093/mnras/stab3048}

\bibitem[{{Smith} \& {MCELS Team}(1999)}]{Smith_1999}
{Smith}, R.~C., \& {MCELS Team}. 1999, in New Views of the Magellanic Clouds,
  ed. Y.~H. {Chu}, N.~{Suntzeff}, J.~{Hesser}, \& D.~{Bohlender}, Vol. 190, 28

\bibitem[{Takahira {et~al.}(2014)Takahira, Tasker, \& Habe}]{Takahira_2014}
Takahira, K., Tasker, E.~J., \& Habe, A. 2014, The Astrophysical Journal, 792,
  63, \dodoi{10.1088/0004-637x/792/1/63}

\bibitem[{Tan {et~al.}(2014)Tan, Beltr{\'{a}}n, Caselli, Fontani, Fuente,
  Krumholz, McKee, \& Stolte}]{Tan_2014}
Tan, J.~C., Beltr{\'{a}}n, M.~T., Caselli, P., {et~al.} 2014, in Protostars and
  Planets {VI} (University of Arizona Press),
  \dodoi{10.2458/azu_uapress_9780816531240-ch007}

\bibitem[{{Tanaka} {et~al.}(2020){Tanaka}, {Zhang}, {Hirota}, {Sakai},
  {Motogi}, {Tomida}, {Tan}, {Rosero}, {Higuchi}, {Ohashi}, {Liu}, \&
  {Sugiyama}}]{TanakaK_2020}
{Tanaka}, K. E.~I., {Zhang}, Y., {Hirota}, T., {et~al.} 2020, \apjl, 900, L2,
  \dodoi{10.3847/2041-8213/abadfc}

\bibitem[{Tokuda {et~al.}(2018)Tokuda, Onishi, Saigo, Matsumoto, Inoue, ichiro
  Inutsuka, Fukui, Machida, Tomida, Hosokawa, Kawamura, \&
  Tachihara}]{Tokuda_2018}
Tokuda, K., Onishi, T., Saigo, K., {et~al.} 2018, The Astrophysical Journal,
  862, 8, \dodoi{10.3847/1538-4357/aac898}

\bibitem[{Tokuda {et~al.}(2019)Tokuda, Fukui, Harada, Saigo, Tachihara, Tsuge,
  Inoue, Torii, Nishimura, Zahorecz, Nayak, Meixner, Minamidani, Kawamura,
  Mizuno, Indebetouw, Sewi{\l}o, Madden, Galametz, Lebouteiller, Chen, \&
  Onishi}]{Tokuda_2019}
Tokuda, K., Fukui, Y., Harada, R., {et~al.} 2019, The Astrophysical Journal,
  886, 15, \dodoi{10.3847/1538-4357/ab48ff}

\bibitem[{{Tokuda} {et~al.}(2020){Tokuda}, {Muraoka}, {Kondo}, {Nishimura},
  {Tosaki}, {Zahorecz}, {Onodera}, {Miura}, {Torii}, {Kuno}, {Fujita}, {Sano},
  {Onishi}, {Saigo}, {Fukui}, {Kawamura}, \& {Tachihara}}]{Tokuda_2020}
{Tokuda}, K., {Muraoka}, K., {Kondo}, H., {et~al.} 2020, \apj, 896, 36,
  \dodoi{10.3847/1538-4357/ab8ad3}

\bibitem[{{Tokuda} {et~al.}(2021){Tokuda}, {Kondo}, {Ohno}, {Konishi}, {Sano},
  {Tsuge}, {Zahorecz}, {Goto}, {Neelamkodan}, {Wong}, {Sewi{\l}o}, {Fukushima},
  {Takekoshi}, {Muraoka}, {Kawamura}, {Tachihara}, {Fukui}, \&
  {Onishi}}]{Tokuda_2021}
{Tokuda}, K., {Kondo}, H., {Ohno}, T., {et~al.} 2021, \apj, 922, 171,
  \dodoi{10.3847/1538-4357/ac1ff4}

\bibitem[{{Torii} {et~al.}(2021){Torii}, {Tokuda}, {Tachihara}, {Onishi}, \&
  {Fukui}}]{Torii_2021}
{Torii}, K., {Tokuda}, K., {Tachihara}, K., {Onishi}, T., \& {Fukui}, Y. 2021,
  \pasj, 73, 205, \dodoi{10.1093/pasj/psaa115}

\bibitem[{Torii {et~al.}(2017)Torii, Hattori, Hasegawa, Ohama, Haworth, Shima,
  Habe, Tachihara, Mizuno, Onishi, Mizuno, \& Fukui}]{Torii_2017}
Torii, K., Hattori, Y., Hasegawa, K., {et~al.} 2017, The Astrophysical Journal,
  835, 142, \dodoi{10.3847/1538-4357/835/2/142}

\bibitem[{Tsuge {et~al.}(2019)Tsuge, Sano, Tachihara, Yozin, Bekki, Inoue,
  Mizuno, Kawamura, Onishi, \& Fukui}]{Tsuge_2019}
Tsuge, K., Sano, H., Tachihara, K., {et~al.} 2019, The Astrophysical Journal,
  871, 44, \dodoi{10.3847/1538-4357/aaf4fb}

\bibitem[{{van den Bergh}(1981)}]{van_1981}
{van den Bergh}, S. 1981, \aaps, 46, 79

\bibitem[{{V{\'a}zquez-Semadeni} {et~al.}(2019){V{\'a}zquez-Semadeni}, {Palau},
  {Ballesteros-Paredes}, {G{\'o}mez}, \& {Zamora-Avil{\'e}s}}]{Vazquez_2019}
{V{\'a}zquez-Semadeni}, E., {Palau}, A., {Ballesteros-Paredes}, J.,
  {G{\'o}mez}, G.~C., \& {Zamora-Avil{\'e}s}, M. 2019, \mnras, 490, 3061,
  \dodoi{10.1093/mnras/stz2736}

\bibitem[{Wong {et~al.}(2011)Wong, Hughes, Ott, Muller, Pineda, Bernard, Chu,
  Fukui, Gruendl, Henkel, Kawamura, Klein, Looney, Maddison, Mizuno, Paradis,
  \& Seale}]{Wong_2011}
Wong, T., Hughes, A., Ott, J., {et~al.} 2011, The Astrophysical Journal
  Supplement Series, 197, 16, \dodoi{10.1088/0067-0049/197/2/16}

\bibitem[{{Young} {et~al.}(2004){Young}, {Lee}, {Evans}, {Goldsmith}, \&
  {Doty}}]{Young_2004}
{Young}, K.~E., {Lee}, J.-E., {Evans}, Neal~J., I., {Goldsmith}, P.~F., \&
  {Doty}, S.~D. 2004, \apj, 614, 252, \dodoi{10.1086/423609}

\bibitem[{{Zhang} {et~al.}(2021){Zhang}, {Zavagno}, {L{\'o}pez-Sepulcre},
  {Liu}, {Louvet}, {Figueira}, {Russeil}, {Wu}, {Yuan}, \&
  {Pillai}}]{Zhang_2021}
{Zhang}, S., {Zavagno}, A., {L{\'o}pez-Sepulcre}, A., {et~al.} 2021, \aap, 646,
  A25, \dodoi{10.1051/0004-6361/202038421}

\bibitem[{Zinnecker \& Yorke(2007)}]{Zinnecker_2007}
Zinnecker, H., \& Yorke, H.~W. 2007, Annual Review of Astronomy and
  Astrophysics, 45, 481, \dodoi{10.1146/annurev.astro.44.051905.092549}

\end{thebibliography}

\end{document}